\documentclass[preprint, showpacs,prd]{revtex4}

\usepackage{graphicx} %To include figures
\usepackage{amsmath}
\usepackage{amssymb}
\usepackage{bm}
\newcommand{\hbbn}[2]{\frac{\partial #1}{\partial #2}}

\def\M{\mathcal{M}}
\def\N{\mathcal{N}}
\def\W{\mathcal{W}}
\def\P{\mathcal{P}}

\def\p{\partial}

\begin{document}

\title{Hamiltonian Dynamics of Spatially-Homogeneous Vlasov-Einstein Systems}

\author{Takahide Okabe\footnote{email: takahideokabe@gmail.com}}
\affiliation{Physics Department and Institute for Fusion Studies,
  The University of Texas at Austin, 1 University Station C1600 Austin, TX 78712-0264, USA}

\author{P. J. Morrison\footnote{email: morrison@physics.utexas.edu}}
\affiliation{Physics Department and Institute for Fusion Studies,
  The University of Texas at Austin, 1 University Station C1600 Austin, TX 78712-0264, USA}

\author{J. E. Friedrichsen III\footnote{email: jfriedri@austincc.edu}}
\affiliation{Physics Department, Austin Community College District --- Riverside Campus, 1020 Grove Blvd. Austin, TX 78741-3337, USA}

\author{L. C. Shepley\footnote{email: shepley@physics.utexas.edu}}
\affiliation{Physics Department and Center for Relativity, The University of Texas at Austin, 1 University Station C1600 Austin, TX 78712-0264, USA}

\date{\today}

\pacs{04.20.Fy, 98.80.Jk, 52.25.Dg, 11.10.Ef}

\begin{abstract}

We introduce a new matter action principle, with a wide range of applicability, for the Vlasov equation in terms of a conjugate pair of functions.
Here we apply this action principle to the study of matter in Bianchi cosmological models in general relativity. The Bianchi models are
spatially-homogeneous solutions to the Einstein field equations, classified by the three-dimensional Lie algebra that describes the symmetry group of
the model. The Einstein equations for these models reduce to a set of coupled ordinary differential equations. The class A Bianchi models admit a
Hamiltonian formulation in which the components of the metric tensor and their time derivatives yield the canonical coordinates. The evolution of anisotropy in the vacuum Bianchi
models is determined by a potential due to the curvature of the model, according to its symmetry.  For illustrative purposes, we examine the evolution
of anisotropy in  models with Vlasov matter.  The Vlasov content is further simplified by the assumption of cold, counter-streaming matter, a kind of
matter that is far from thermal equilibrium and is not describable by an ordinary fluid model nor other more simplistic matter models. Qualitative
differences and similarities are found in the dynamics of certain vacuum class A Bianchi models and Bianchi Type I models with cold,
counter-streaming Vlasov matter potentials analogous to the curvature potentials of corresponding vacuum models.

\end{abstract}

\maketitle

%%%%%%%%%%%%%%%%%%%
%%%%%%%%%%%%%%%%%%%

\section{Introduction}
According to observations, our Universe is both spatially homogeneous and isotropic on large scales \cite{SeanCarroll,SeanCarroll2}.  However, in this
paper a more general case is considered in which the universe is taken to be spatially homogeneous but anisotropic \cite{Wald,RyanShepley}. We further
assume that our models contain matter governed by the Vlasov equation \cite{Stewart,EhlersKineticTheory, BerezSachs}, which opens the possibility that
it can be anisotropic in momentum space as well.  Spatially-homogeneous cosmological models, which are also known as Bianchi cosmologies 
\cite{RyanShepley}, are classified into nine types based on a standard classification of three-dimensional Lie algebras \cite{Taub} that determine the
symmetry of the model. 

The presence of matter influences the dynamics of the model. In this paper we study Vlasov matter: collisionless particles interacting only through
their mutual gravitational effect. We introduce a new action principle for such matter, applicable in a wide variety of contexts.  Hamiltonian methods 
have long had important roles in general relativity \cite{Dirac,ADM}.  In a model
that is filled with Vlasov matter in a spatially-homogeneous way, the presence of matter appears as an additional potential term in the Hamiltonian
\cite{TakahideDissertation}. These additional potentials are not as steep as the curvature potentials of the vacuum Bianchi cosmologies, so the
evolution of anisotropy is always under the influence of the matter potential, in contrast to the vacuum cases \cite{TakahideDissertation}. In this
paper, vacuum Bianchi cosmologies and Type I models with Vlasov matter are compared in order to clarify the distinction between similarly shaped
curvature and matter potentials.

There are two main reasons for studying Bianchi cosmologies. First, universes that are spatially homogeneous but anisotropic are the simplest
generalization of the spatially-homogeneous and isotropic universes, because there are no models that are everywhere isotropic and spatially
inhomogeneous \cite{Wald}.  From the theoretical point of view, it is of interest to see how the dynamics of a model universe is changed by this first
step away from the spatially isotropic cosmologies.  Second, due to the spatial homogeneity, the Einstein equations are reduced to a set of coupled
ordinary differential equations, which can be viewed as a finite-dimensional dynamical system.  

A reason for studying matter described by the Vlasov equation is that this model allows for phase space degrees of freedom.  Consequently,  with it
the dynamics of nonthermalized matter can be explored.  The reduction to a set of ordinary differential equations is also possible with this kind of
matter description within the Bianchi cosmology context. We believe that physically meaningful cosmological models should have realistic matter
models.  Although both the Bianchi cosmologies and the Vlasov matter we consider are simplistic, they are less so than the common practice of merely
describing matter dynamics by a perfect fluid.  We thus agree with several others in this goal (see \cite{Andreasson}).

A spatially-homogeneous spacetime is foliated by spacelike hypersurfaces, on which the metric tensor $\mathbf{g}$ is invariant under the transitive
action of an isometry group.  If the group is a simply transitive Lie group (and thus three-dimensional), symmetry is characterized by three
linearly-independent, spatial Killing vector fields $\{\boldsymbol{\xi}_{(i)}\}_{i=1}^3$.  Their  constant structure coefficients $C^k_{\ ij}$
determine the nature of the symmetry: 
\begin{equation} [\boldsymbol{\xi}_{(i)},\boldsymbol{\xi}_{(j)}]=C^{k}_{\ ij}\boldsymbol{\xi}_{(k)}\,,
\end{equation} 
where the Einstein summation convention is used, as it will be throughout the paper.  The three-dimensional Lie groups (actually their
Lie algebras) are classified into nine Bianchi types.  This classification of the symmetry of the spacelike hypersurfaces defines the Bianchi
cosmological models, which were further subclassified as class A and class B models by Ellis and MacCallum \cite{MacCallum3}.  The models of interest
in this paper, the class A models, are characterized by $C^k_{\ ik}=0$. The structure coefficients used in this paper for the class A models are in
common standard forms that are summarized in Table \ref{table: StructureConstantsClassA}.

%%%%%%%%%%%%%%%%%%%%%%%%%%%%%%%%%%%%%%%%%%%%%%%%%%%%%%%%%%%%%%%%
\begin{table}[htbp]
\caption{The structure coefficients 
(structure constants)
of class A Bianchi models \cite{RyanShepley}.  
These coefficients are given in standard form and of course may vary under linear transformations of the basis vectors.} 
\begin{ruledtabular}
\begin{tabular}{rl}
\textbf{Type}\ & \textbf{Structure Constants}\\
\hline
\textbf{I} &  $C^{k}_{\ ij}=0$ \\
\textbf{II} & $C^{1}_{\ 23}=-C^{1}_{\ 32}=1$ \quad (other: $0$)\\
\textbf{VI}$_{\bm{0}}$ & $C^{1}_{\ 23}=-C^{1}_{\ 32}=1$, $C^2_{\ 13}=-C^2_{\ 31}=1$ \quad (other: $0$)\\
\textbf{VII}$_{\bm{0}}$ & $C^{1}_{\ 23}=-C^{1}_{\ 32}=-1$, $C^{2}_{\ 13}=-C^{2}_{\ 31}=1$ \quad (other: $0$)\\
\textbf{VIII} & $C^{1}_{\ 23}=-C^{1}_{\ 32}=-1$, $C^{2}_{\ 31}=-C^{2}_{\ 13}=1$, $C^{3}_{\ 12}=-C^{3}_{\ 21}=1$ \quad (other: $0$)\\
\textbf{IX} & $C^{k}_{\ ij}=\epsilon^{ijk}$  \quad  (the Levi-Civita symbol)
\end{tabular}
\end{ruledtabular}
\label{table: StructureConstantsClassA}
\end{table}
%%%%%%%%%%%%%%%%%%%%%%%%%%%%%%%%%%%%%%%%%%%%%%%%%%%%%%%%%%%%%%%%

When a class A Bianchi-type symmetry is imposed on the vacuum Einstein equations, the equations become a coupled set of ordinary differential
equations that can be written in Hamiltonian form.  The potential terms arise from the curvature of the model which is dictated by the symmetry.  The
curvature potential terms in the various Hamiltonians (see Table \ref{table: CurvaturePotentials}) result in qualitatively distinct evolutions of
anisotropy in the vacuum models \cite{RyanShepley}. In order to discuss the influences of the curvature potentials, it is necessary to understand the
dynamics of the model universe in the absence of the curvature potential terms, namely the vacuum Bianchi Type I model.  For this  model  the corresponding Hamiltonian is integrable, and the explicit solutions are known as the Kasner solutions. In the other Bianchi models the presence of a
curvature potential affects the dynamics.  The Kasner solutions are important because the dynamics of the other models can be approximated as a series
of Kasner solutions with different Kasner parameters \cite{RyanShepley}.  The curvature potentials are so steep that they can be approximated as moving
potential walls within which the point that represents the state of the universe moves as an approximate Kasner solution whose parameters change when
the universe point bounces off the walls. This approximation is known as qualitative cosmology \cite{RyanDissertation,CollinsStewart,Collins,
RyanShepley}.

The paper is organized as follows:  In the next section, the Hamiltonian approach to Vlasov matter, based on a pair of conjugate potential functions,
is presented in a general context.  This variational principle, although based on earlier works  \cite{morrison81,morrison82,MorrisonYe},  is new.  Spatially-homogeneous universes are discussed in a general way, and the Vlasov Hamiltonian is given
in a form compatible with the symmetry. In the ensuing section, the evolution of anisotropy in the presence of Vlasov matter is analyzed by comparing
and contrasting certain vacuum Bianchi models with those having Vlasov matter in a Type I model.   (A cold, counter-streaming distribution function
that supports the spatial symmetry is assumed.)  The last section is devoted to conclusions and suggestions for further research.

%%%%%%%%%%%%%%%%
%%%%%%%%%%%%%%%%

\section{Formulation}

\subsection{Variational Principle and Derivation of Hamiltonian}

In Vlasov theory \cite{KandrupNeill,MorrisonYe,Morrison80,KandrupMorrison}, matter is modeled by a phase-space distribution function denoted by $F(x^\mu, p_\nu)$, where the $\{x^\mu \}$ are the positions of the particles and the $\{ p_\nu \}$ are
their 4-momenta.  The systems that are studied here consist of particles of the same mass, so that  a mass-shell constraint will be imposed:
\begin{equation}
g^{\mu \nu}p_{\mu} p_{\nu}=-m^{2} \quad {\mathrm{and}} \quad p^{0}>0. \label{eq: mass-shell}
\end{equation}
For the purposes of this exposition, we set a gauge condition on the spacetime metric $g_{\mu\nu}$ such that  $g_{00} = -1$ and $g_{0i} = 0$ for $i =
1, 2, 3$ (we later generalize this gauge). The invariant volume element in 4-momentum space is reduced to the following volume element on the
3-momentum mass-shell, with components $p_i$: 
\begin{eqnarray}
\frac{d^{3}p}{\sqrt{-g}\, \bar{p}^{0}},\hspace{10 pt}{\mathrm{with}}\hspace{10 pt} \bar{p}^0 := \sqrt{m^2+ g^{ij}p_ip_j}
\quad {\rm and}\quad g:=\det|g_{\mu\nu}|. \label{eq: p-bar-0}
\end{eqnarray}
Because of the gauge choice and the mass-shell constraint, it is convenient to define an on-shell distribution function,
\begin{eqnarray}
f(x^{\mu}, p_{i}):=F \bigl( x^{\mu}, p_{i}, \bar{p}_{0}(x^{\mu}, p_{i}) \bigr),
\end{eqnarray}
where $\bar{p}_0 =- \bar{p}^0$ is the momentum defined in equation (\ref{eq: p-bar-0}).  The notation $\bar{p}_{\mu}$ will be used to indicate the three variables
$\bar{p}_i = p_i$ along with the functional form of $\bar{p}_0 = - \bar{p}^0$.

Since matter is assumed to be collisionless, $F$ is constant along geodesics, and so $F$ and $f$ are governed by the relevant Vlasov equation. The
off-shell Vlasov equation is simply the geodesic equation for the particle paths (the momentum one-form $\mathbf{p}$ is the mass $m$ times the
particle velocity):
\begin{equation}
p^{\mu}\frac{\partial F}{\partial x^{\mu}}-\frac{1}{2}g^{\alpha \beta}_{\ \ ,\mu}\,p_{\alpha}p_{\beta}\frac{\partial F}{\partial p_{\mu}}=0.
\end{equation}
The on-shell Vlasov equation is obtained by using equation (\ref{eq: mass-shell}). It can be expressed as
\begin{equation}
 \frac{\partial f}{ \partial t}+\left\{f,  \bar{p}^{0}\right\}_{3}=0, \label{eq: Vlasov}
\end{equation}
where $\{\cdot, \cdot\}_3$ is the three-dimensional Poisson bracket:
\begin{equation}
\left\{f,  \bar{p}^{0}\right\}_{3}=\frac{1}{\bar{p}^0}
\left(\bar{p}^i\hbbn{f}{x^i}-\frac{1}{2}g^{\alpha \beta}_{\ \ ,i}\,\bar{p}_{\alpha}\bar{p}_{\beta}\frac{\partial f}{\partial p_{i}} \right)\,.
\end{equation}

The metric tensor $\mathbf{g}$ in the Vlasov equation is determined by the Einstein equations (where the coupling constant is taken to be unity):
\begin{equation}
R_{\mu \nu}-\frac{1}{2}R g_{\mu \nu}+\Lambda g_{\mu \nu}= 
T_{\mu \nu}:=
\int \frac{d^3p}{\sqrt{-g}\, \bar{p}^0} \bar{p}_\mu \bar{p}_\nu  f(x^{\mu}, p_i)\,.
 \label{eq: Einstein}
\end{equation}
The action functional that gives equation (\ref{eq: Einstein}) is
\begin{equation}
S=S_{{\mathrm{Hilbert}}}+S_{\Lambda}+S_{{\mathrm{matter}}}=\int d^{4}x\, \sqrt{-g} R-2\Lambda\int d^{4}x\, \sqrt{-g}
+S_{\mathrm{matter}}.
\end{equation}
The Hilbert action $S_{\mathrm{Hilbert}}$ gives the Einstein tensor and $S_\Lambda$ gives the cosmological term upon variation with respect to
$g^{\mu\nu}$ and when necessary integrating by parts and ignoring the boundary terms \cite{SeanCarroll,SeanCarroll2}:
\begin{equation}
\frac{1}{\sqrt{-g}}\frac{\delta S_{\mathrm{Hilbert}}}{\delta g^{\mu\nu}}=R_{\mu\nu}-\frac{1}{2}Rg_{\mu\nu}\quad {\mathrm{and}} \quad \frac{1}{\sqrt{-g}}\frac{\delta S_{\Lambda}}{\delta g^{\mu\nu}}=\Lambda g_{\mu\nu}.
\end{equation}
$S_{\mathrm{matter}}$ must satisfy
\begin{equation}
\frac{1}{\sqrt{-g}}\frac{\delta S_{\mathrm{matter}}}{\delta g^{\mu\nu}}=-T_{\mu\nu}.
\end{equation}
An action using $S_{\mathrm{matter}}$ must also yield the Vlasov equation.

In order to put the Vlasov equation into a variational form, we express the distribution function  
by a conjugate pair of phase-space functions $(\M, \N)$,  using the four-dimensional Poisson bracket $\{\cdot,\cdot \}_4$  \cite{morrison81,morrison82,MorrisonYe}: 
\begin{equation}
F(x^{\mu}, p_{\nu})=\{\M, \N\}_{4}:=
\frac{\partial \M}{\partial x^{\mu}} \frac{\partial \N}{\partial p_{\mu}}
-\frac{\partial \M}{\partial p_{\mu}}\frac{\partial \N}{\partial x^{\mu}}. 
\end{equation}
$\M(x^{\mu}, p_{\nu})$ and $\N(x^{\mu}, p_{\nu})$ are each required to satisfy Vlasov equations, which are compactly written as
\begin{eqnarray}
\left\{
\left(
\begin{array}{c}
\M\\
\N\\
\end{array}
\right),\,  
\frac{1}{2}g^{\mu \nu}p_{\mu}p_{\nu} \right\}_{4}=0.
\label{Equation_for_AB}
\end{eqnarray}
The Vlasov equation for $F(x^{\mu}, p_{\nu})$ \cite{MMMT86},  
\begin{equation}
\Bigr\{ F ,\, \frac{1}{2}g^{\mu\nu}p_{\mu}p_{\nu} \Bigr\}_4=
\Bigr\{ \{\M,\N\}_4, \, \frac{1}{2}g^{\mu\nu}p_{\mu}p_{\nu} \Bigr\}_4=0\,, 
\end{equation}
is derived by making use of the Jacobi identity \cite{morrison82,MorrisonYe}. The on-shell restriction can be achieved by defining
\begin{eqnarray}
\mu(x^{\mu}, p_{i}):=\M(x^{\mu}, p_{i}; \bar{p}_{0}(x^{\mu}, p_{i})), \quad 
\nu(x^{\mu}, p_{i}):=\N(x^{\mu}, p_{i}; \bar{p}_{0}(x^{\mu}, p_{i})). 
\end{eqnarray}
Therefore, the on-shell distribution function is expressed by the on-shell pair: \begin{equation}
 f(x^{\mu}, p_i)=\{\mu, \nu\}_3. 
 \label{mu-nu-distribution}
\end{equation}
The equations that $\mu$ and $\nu$ have to satisfy are also derived by imposing the mass-shell condition on equations (\ref{Equation_for_AB}), 
\begin{equation}
 \frac{\partial}{\partial t}\left(
\begin{array}{c}
\mu\\
\nu\\
\end{array}
\right)
+\left\{ \left(
\begin{array}{c}
\mu\\
\nu\\
\end{array}
\right),\, \bar{p}^{0}\right\}_{3}=0, \label{eq: var210}
\end{equation}
from which the on-shell Vlasov equation for $f(x^{\mu}, p_i)$ is derived:
\begin{equation}
\hbbn{}{t}\{\mu, \nu \}_3+ \Bigr\{ \{\mu, \nu\}_3, \, \bar{p}^0 \Bigr\}_3=0\,.
\end{equation}
Again, the derivation rests on the Jacobi identity.

There is a gauge group associated with the introduction of the pairs $(\M,\N)$ or $(\mu,\nu)$.  For example,  a transformation of the form
$(\bar{\mu}(\mu, \nu),  \bar{\nu}(\mu, \nu))$ that satisfies the Jacobian condition $\p(\bar\mu, \bar\nu)/\p(\mu, \nu)=1$, ensures that $ \{\bar{\mu},
\bar\nu\}_3= \{\mu, \nu\}_3$.  A full discussion of the gauge group associated with this `potential'  decomposition, including both dependent and
independent variables,  will be presented in a subsequent paper. 

Now that the distribution function is expressed by the introduction of a conjugate pair of phase space functions, the matter action for the
Vlasov-Einstein system can be written down. The action principle in terms of these variables is canonical, that is the evolution of those variables is
manifestly determined by a canonical Poisson bracket.  We check the variational formulations in both off-shell and on-shell forms and describe how
they are naturally related to each other. 

If there is no mass-shell constraint, the matter action is simply
\begin{equation}
S_{{\mathrm{off}}}[\M, \N, g^{\mu \nu}]= -\int d^4 x\ d^4 p\, g^{\rho \sigma}p_{\rho}p_{\sigma}\{\M, \N\}_{4},  \label{eq: offShellAction}
\end{equation}
and the functional derivatives are
\begin{equation}
\frac{1}{\sqrt{-g}}\frac{\delta S_{{\mathrm{off}}}}{\delta g^{\mu \nu}}= - \int \frac{d^{4}p}{\sqrt{-g}} p_{\mu}p_{\nu} \{\M, \N\}_{4}=-T_{\mu\nu}, \label{eq: var150}
\end{equation}
and
\begin{equation}
\left. \begin{array}{r@{\quad =\quad}l} \delta S_{\mathrm{off}}/\delta \M& 0 \\ \delta S_{\mathrm{off}}/\delta \N& 0 \end{array} \right\} \Leftrightarrow \left\{ \begin{array}{r@{\quad =\quad}l} \left( p^{\mu}\frac{\partial }{\partial x^{\mu}}-\frac{1}{2}g^{\mu \nu}_{\ \ ,\rho}\ p_{\mu}p_{\nu}\frac{\partial }{\partial p_{\rho}} \right)\N & 0,  \\ \left( p^{\mu}\frac{\partial }{\partial x^{\mu}}-\frac{1}{2}g^{\mu \nu}_{\ \ ,\rho}\ p_{\mu}p_{\nu}\frac{\partial }{\partial p_{\rho}} \right)\M & 0. \end{array} \right.
\end{equation}

Since we consider a system of equal mass particles, (\ref{eq: offShellAction}) has to be put on-shell.  This can be done by restricting $\M$ and
$\N$ to be on-shell and integrating (\ref{eq: offShellAction}) over $p_0$, either before or after the variation.  (In other words the on-shell
constraint and the variation commute.)  The integration over $p_0$ is best achieved with the use of the function \cite{AndersonSpiegel}
\begin{equation}
\delta_{+}(-g^{\mu \nu}p_{\mu}p_{\nu}-m^2):=2\Theta(p^{0})\delta(-g^{\mu \nu}p_{\mu}p_{\nu}-m^2)
=\frac{1}{\bar{p}^0}\delta(p_{0}-\bar{p}_{0}). \label{eq: constraintDelta}
\end{equation}

However, care should be taken with the independent dynamical variables.  If variations are done before the on-shell constraint, then the independent
dynamical variables are of course $g^{\mu \nu}, \M,$ and $\N$, but if variations are done after the on-shell constraint, they are $g^{\mu \nu}, \mu,$
and $\nu$.  If the variational derivatives are taken before the on-shell constraint, the functional derivatives are found to be
\begin{equation}
\frac{1}{\sqrt{-g}}\frac{\delta S}{\delta g^{\mu \nu}}= -\int \frac{d^{4}p}{\sqrt{-g}} p_{\mu}p_{\nu} \underbrace{\{\M,\N\}_{4}}_{=: F(x^{\mu}, p_{\nu})} \delta_{+}({\mathrm{on\ shell}})=-T_{\mu \nu}, \label{eq: var190}
\end{equation}
and
\begin{equation}
\left. \begin{array}{r@{\quad =\quad}l} \delta S/\delta \M& 0 \\ \delta S/\delta \N& 0 \end{array} \right\} \Leftrightarrow \left\{ \begin{array}{r@{\quad =\quad}l} \left( \left[ p^{\mu}\frac{\partial }{\partial x^{\mu}}-\frac{1}{2}g^{\mu \nu}_{\ \ ,\rho}\ p_{\mu}p_{\nu}\frac{\partial }{\partial p_{\rho}} \right]\N \right)\delta_{+}({\mathrm{on\ shell}}) & 0, \\ \left( \left[ p^{\mu}\frac{\partial }{\partial x^{\mu}}-\frac{1}{2}g^{\mu \nu}_{\ \ ,\rho}\ p_{\mu}p_{\nu}\frac{\partial }{\partial p_{\rho}} \right]\M \right)\delta_{+}({\mathrm{on\ shell}}) & 0. \end{array} \right.
 \label{eq: var200}
\end{equation}
The on-shell delta functions in (\ref{eq: var200}) can be eliminated by integrating them with respect to $p_{0}$, which results in (\ref{eq: var210}).
 The calculation of (\ref{eq: var190}) is not as straightforward as (\ref{eq: var150}), because the $g^{\mu \nu}$ in $\delta_{+}({\mathrm{on\
shell}})$ is subject to the variation.  The calculation is lengthy, but all the extra terms vanish to give (\ref{eq: var190})
\cite{TakahideDissertation}.

On the other hand, the mass-shell constraint can be imposed before variations.  For this case, the action is given by 
\begin{equation}
S_{{\mathrm{on}}}[\mu, \nu, g^{\mu\nu}]
= 2 \int d^{4}x\ d^{3}p \left( \nu\frac{\partial \mu}{\partial t}- \bar{p}^{0} \{\mu, \nu\}_{3} \right),
 \label{eq: var201}
\end{equation}
whose functional derivatives are
\begin{equation}
\frac{1}{\sqrt{-g}}\frac{\delta S_{{\mathrm{on}}}}{\delta g^{\mu \nu}}
= - \int \frac{d^{3}p}{\sqrt{-g}\, \bar{p}^{0}} \, \bar{p}_{\mu} \bar{p}_{\nu} \{\mu, \nu\}_{3}=- T_{\mu \nu}, 
\end{equation}
and
\begin{equation}
\left. \begin{array}{r@{\quad =\quad}l} \delta S_{\mathrm{on}}/\delta \mu& 0 
\\ 
\delta S_{\mathrm{on}}/\delta \nu& 0 \end{array}\right\}  \Leftrightarrow \left\{ \begin{array}{r@{\quad =\quad}l} \partial \nu/ \partial t+\{\nu, \bar{p}^0\}_3 & 0, \\ \partial \mu/\partial t+\{\mu, \bar{p}^0\}_3  & 0.
 \end{array} 
 \right.
 \label{munu}
\end{equation}
Note that the integrand of (\ref{eq: var201}) is reminiscent of the phase space action, ``$p\dot{q}-H$", since $\mu$ and $\nu$ are canonically
conjugate to each other and $\bar{p}^0=-\bar{p}_{0}$ is the energy of a particle with $\{\mu, \nu\}_{3}$ being the particle distribution on phase
space.  Consequently (\ref{munu})  have the Hamiltonian form:
\begin{equation}
\frac{\p \mu}{\p t}= \frac{\delta H}{\delta \nu}
\qquad {\rm and}\qquad
\frac{\p \nu}{\p t}=- \frac{\delta H}{\delta \mu}\,.
\end{equation}

%%%%%%%%%%%%%%%%%%%%%
%%%%%%%%%%%%%%%%%%%%%

\subsection{Vlasov Matter in a Spatially-Homogeneous Spacetime}

Since the spacetime structure and the matter configuration are related by the Einstein equations, the matter term is subject to symmetry conditions.
If some spacetime symmetry is assumed, the form of a distribution function is restricted so that its corresponding stress-energy tensor has the same
symmetry as the spacetime.  It is sufficient, in the presence of a Killing vector field $\xi^{\mu}$, that \cite{MaartensMaharaj1985}
\begin{equation}
\frac{\partial F}{\partial x^{\mu}}\xi^{\mu}-\frac{\partial F}{\partial p_{\mu}}\xi^{\nu}_{\ ,\mu}p_{\nu}=0.\label{eq: symmetry_condition}
\end{equation}

The quantity
\begin{equation}
Y:=\mathbf{p}(\boldsymbol{\xi})
\end{equation} 
is a constant along the geodesic particle paths. If the Killing vector is spatial, equation (\ref{eq: symmetry_condition}) becomes
\begin{equation}
\{f, Y\}_3=\{f, \mathbf{p}(\boldsymbol{\xi})\}_3=0. \label{eq: on-shell_symmetry}
\end{equation}
Under spatial homogeneity, it is natural to use a left-invariant vector basis $\{\mathbf{X}_{(\mu)}\}$ and its dual basis
$\{\boldsymbol{\sigma}^{(\mu)}\}$ \cite{RyanShepley}. The vector basis is chosen so that the spatial $\{\mathbf{X}_{(i)}\}$ are tangent to the
spatially-homogeneous sections, and $\{\mathbf{X}_{(0)}\}$ is perpendicular to them.  The vector basis satisfies the commutation relations
\begin{eqnarray}
\left[\mathbf{X}_{(0)},\mathbf{X}_{(i)}\right] &=& 0,\\
\left[\mathbf{X}_{(i)},\mathbf{X}_{(j)}\right] &=& -C^s_{\ ij}\mathbf{X}_{(s)}.
\end{eqnarray}
Similarly, the dual basis (with $\boldsymbol{\sigma}^{(0)} = \mathbf{d}t$) satisfies
\begin{eqnarray}
\mathbf{d}\boldsymbol{\sigma}^{(0)} &=& 0 \\
\mathbf{d}\boldsymbol{\sigma}^{(i)}&=& \frac{1}{2} C^i_{\ st}\boldsymbol{\sigma}^{(s)} \wedge \boldsymbol{\sigma}^{(t)}.
\end{eqnarray}
The components of the metric tensor in this basis are functions only of $t$:
\begin{equation}
g_{\mu\nu} = \mathbf{g}\left(\mathbf{X}_{(\mu)},\mathbf{X}_{(\nu)}\right) = g_{\mu\nu}(t).
\end{equation}

Momentum is expressed as 
\begin{equation}
h_{\mu}=\mathbf{\bar{p}}(\mathbf{X}_{(\mu)}) \hspace{10 pt}{\mathrm{i.e.,}}\hspace{3 pt}\mathbf{\bar{p}}=h_{\mu}\boldsymbol{\sigma}^{(\mu)}.
\end{equation}
Note, however, that the $h_{\mu}$ are not canonical coordinates in general, because 
\begin{equation}
\{h_i, h_j\}_3=C^l_{\ ij}h_l\,,
\end{equation}
 while canonical coordinates would satisfy $\{p_i, p_j\}_3=0$.  With the momentum coordinates $h_i$ and the time variable $t$, the
 spatially-homogeneous distribution function has the form,
\begin{equation}
f=f(t, h_i), \label{eq: spatially-homogeneous_dist_func}
\end{equation}
so that $f$ does not depend explicitly on the spatial coordinates. The first term of (\ref{eq: var201}) gives zero upon variation with respect to
$g^{\mu\nu}$, and $\{\mu,\nu\}_3$ can be replaced by $f$. The Vlasov equation can therefore be rewritten, reflecting the symmetry, as follows:
\begin{equation}
\hbbn{f}{t}+\hbbn{f}{h_a}\,C^d_{\ ab}h_d\hbbn{\bar{h}^0}{h_b}=0, \hspace{10 pt} {\mathrm{with}}\hspace{10 pt}  \bar{h}^0 =-\bar{h}_0 =\sqrt{m^2+ g^{ij}h_ih_j}\,.
\end{equation}
It is immediately seen that in Type I (where $C^c_{\ ab}=0$), the solution of  the Vlasov equation is independent of time:  $f=f(h_1,h_2, h_3)$.  

Based on the matter action we derived in the previous sub-section, the Hamiltonians for the class A Bianchi cosmologies with Vlasov matter are
obtained.  The calculation is simplified by restricting to the diagonal case:   The positive-definite spatial metric $g_{ij}$ will be taken to be
diagonal in a basis that is invariant under the symmetry transformation described by the Lie group of the particular Bianchi model
\cite{JamesDissertation}. This is not a gauge choice; it imposes physical limitations on the model (see \cite{MacCallum2, MacCallum3, MacCallum4}) not
only because the standard form for the structure coefficients of the Lie Algebra are used, but also because of restrictions which must be placed on
the particle momenta (as described below).  At this point it is necessary to keep all diagonal components of the metric as variables (rather than
setting $g_{00}=-1$ as we had done earlier) in order to formulate explicit Hamiltonians:
\begin{equation}
g_{\mu \nu}=\mathbf{g}(\mathbf{X}_{(\mu)}, \mathbf{X}_{(\nu)})={\mathrm{diag}}\left(-N^{2}(t), A^{2}(t), B^{2}(t), C^{2}(t)\right). \label{eq: DiagonalMetric}
\end{equation}
As we said, this restriction to the diagonal case, when the standard forms from Table \ref{table: StructureConstantsClassA} are used for the $C^i_{\
jk}$, is really a restriction in a sense that some physical effects are then not allowed.  The $T_{\mu\nu}$ based on the distribution function must also be diagonal; for example,  it is sufficient that the distribution function be an even function of $h_1, h_2,$ and $h_3$.   After obtaining the Hamiltonian, we can specify the form of $N$, by choosing
an appropriate time coordinate.  Thus, the Hamiltonian systems we obtain in this section are three-degree-of-freedom systems.  Note that the
components of the metric tensor depend only on $t$ due to the use of the invariant basis.

The action functional is
\begin{eqnarray}
S_{{\mathrm{Hilbert}}}&+&S_{\Lambda}+S_{{\mathrm{matter}}}=\int dt\, \left[ -\frac{2}{N} \left(\dot{A}\dot{B}C+\dot{A}B\dot{C}+A\dot{B}\dot{C}\right) + ({\mathrm{terms\ with\ }} C^{i}_{\ jk}) \right] \nonumber\\
& &-2\Lambda \int dt\, NABC  -2 \int dt\, \int d^3h\,  N  f(t, h_k) \sqrt{m^2+\frac{h_1^2}{A^2}+\frac{h_2^2}{B^2}+\frac{h_3^2}{C^2}}\,, 
\end{eqnarray}
where $\dot{}$ is $d/dt$ and where ``terms with $C^{i}_{\ jk}$'' appear as a curvature potential term $V_c$ in the Hamiltonian.  By performing the
Legendre transform on the Lagrangian, the Hamiltonian is obtained.  Since there is no $\dot{N}$, $N$ is treated as a Lagrangian multiplier, and the
variation with respect to $N$ gives the Hamiltonian constraint:
\begin{eqnarray}
0 \equiv  H := \frac{N}{8ABC} \left(A^2 \pi_{A}^2+B^2 \pi_{B}^2+C^2 \pi_{C}^2-2AB\pi_{A}\pi_{B}-2BC\pi_{B}\pi_{C}-2CA\pi_{C}\pi_{A} \right) \nonumber\\
   + NV_c(A,B,C)+2N\Lambda ABC+ 2N \int d^{3}h\, f(t, h_k)  \sqrt{m^{2}+\frac{h_{1}^{\ 2}}{A^2}
 +\frac{h_{2}^{\ 2}}{B^2}+\frac{h_{3}^{\ 2}}{C^2}}\,. 
 \label{eq: HamiltonianABC}
\end{eqnarray}

\begin{figure}[htbp]
\begin{tabular}{ccc}
\begin{minipage}{0.3\hsize}
  \begin{center}
    \includegraphics[width=1.0\hsize]{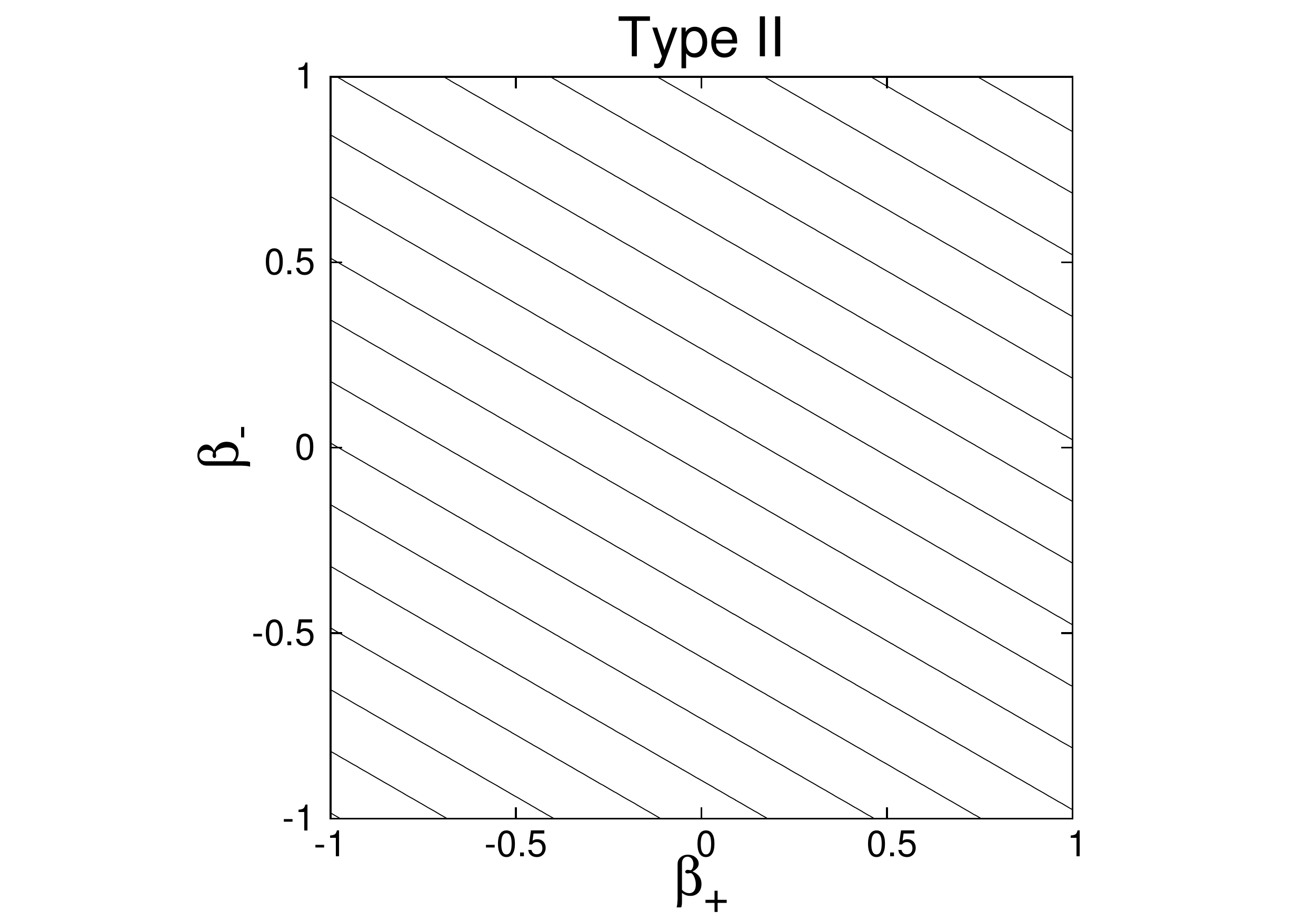}
  \end{center}
\end{minipage}
\begin{minipage}{0.3\hsize}
  \begin{center}
    \includegraphics[width=1.0\hsize]{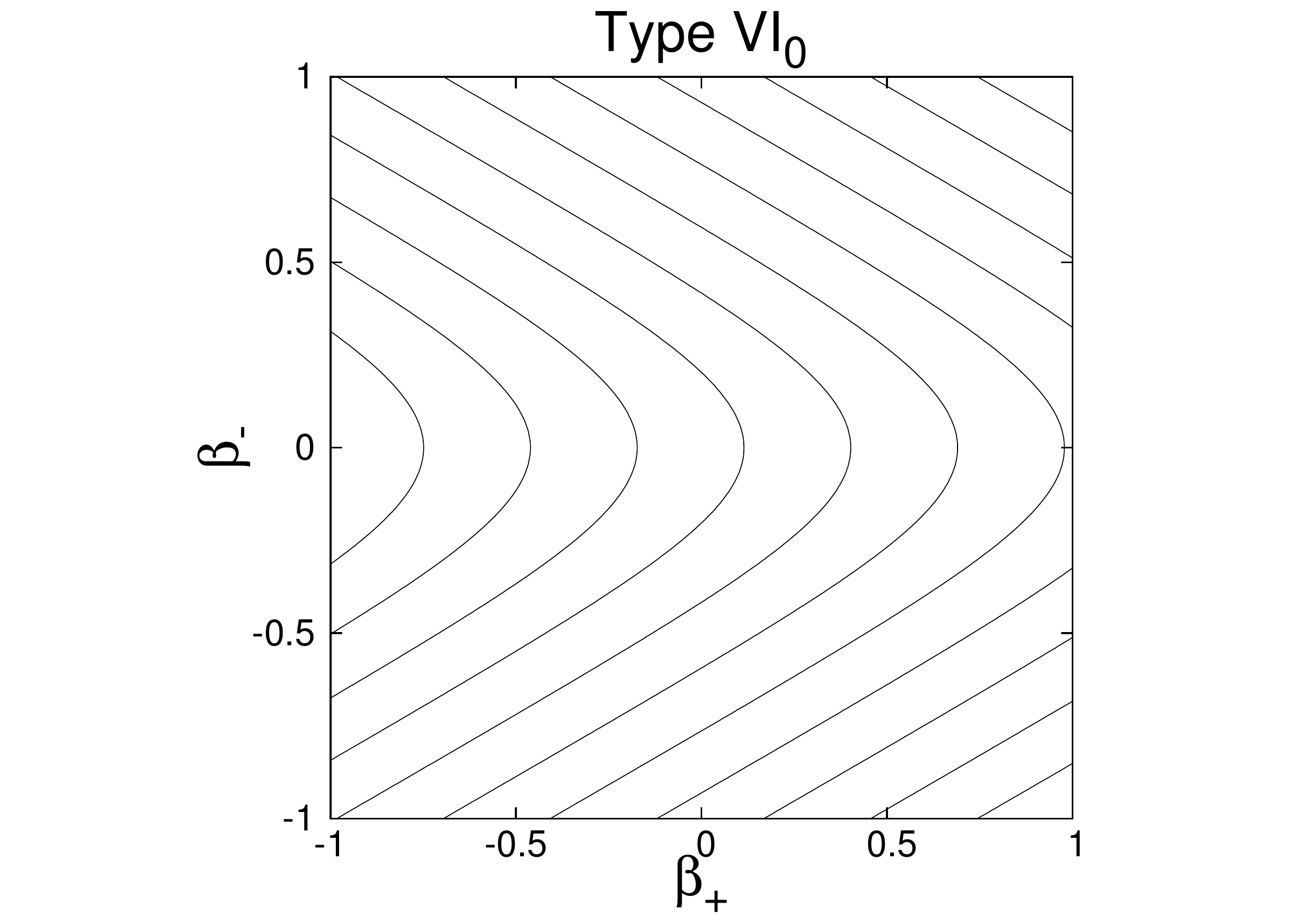}
  \end{center}
\end{minipage}
\begin{minipage}{0.3\hsize}
  \begin{center}
    \includegraphics[width=1.0\hsize]{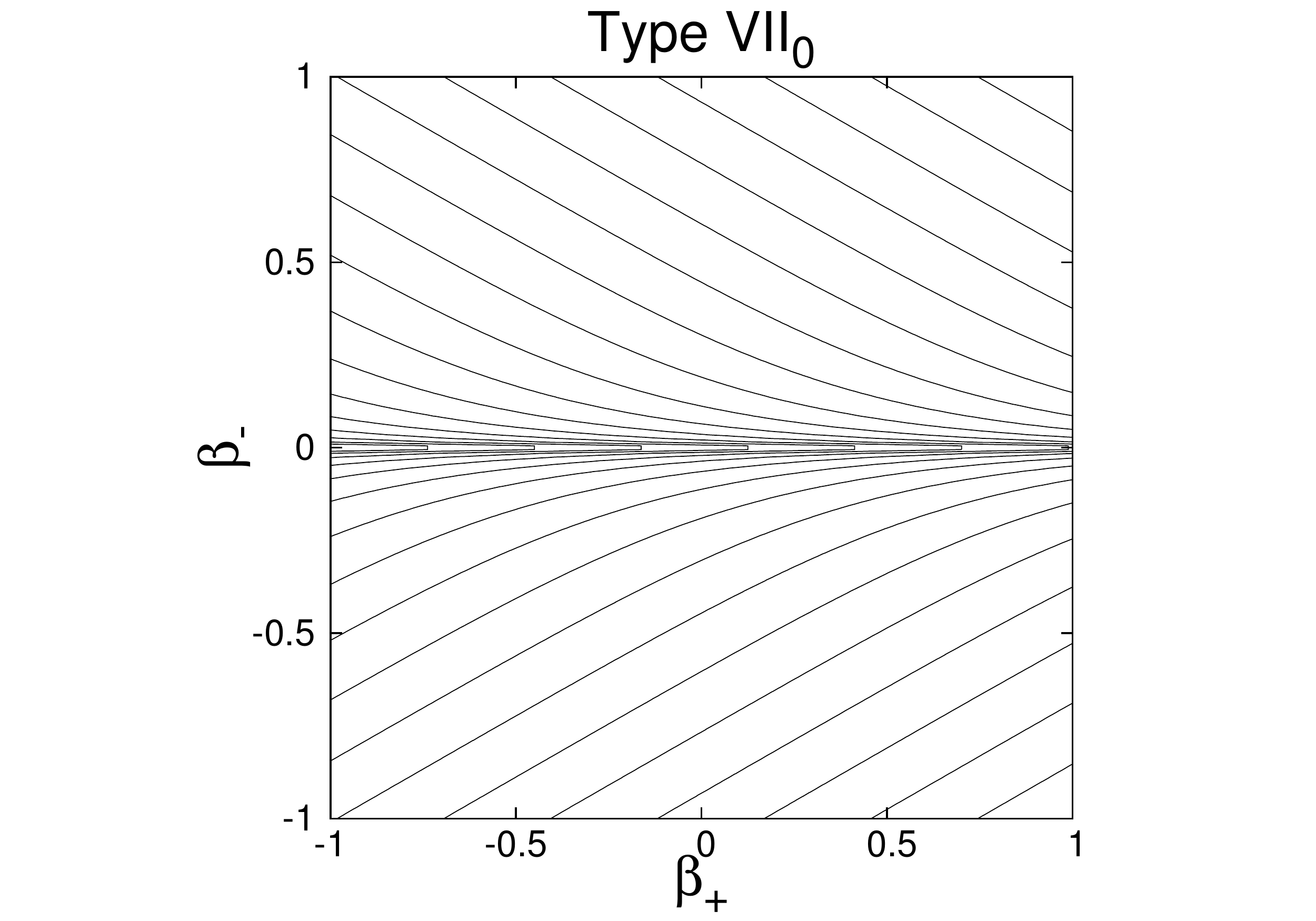}
  \end{center}
\end{minipage}
\end{tabular}
\vspace{1.0 cm}
\begin{tabular}{cc}
\begin{minipage}{0.3\hsize}
  \begin{center}
    \includegraphics[width=1.0\hsize]{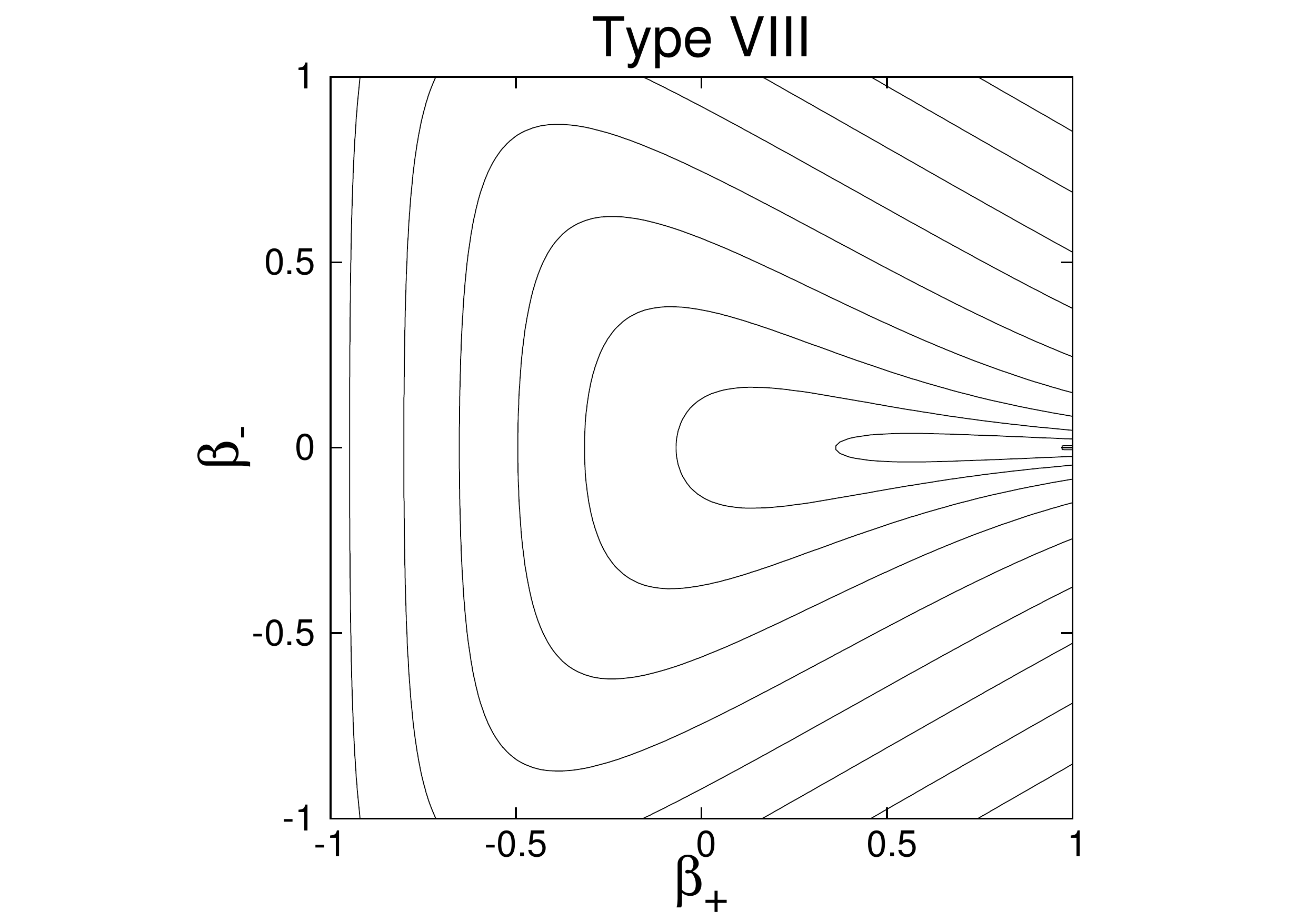}
  \end{center}
\end{minipage}
\begin{minipage}{0.3\hsize}
  \begin{center}
    \includegraphics[width=1.0\hsize]{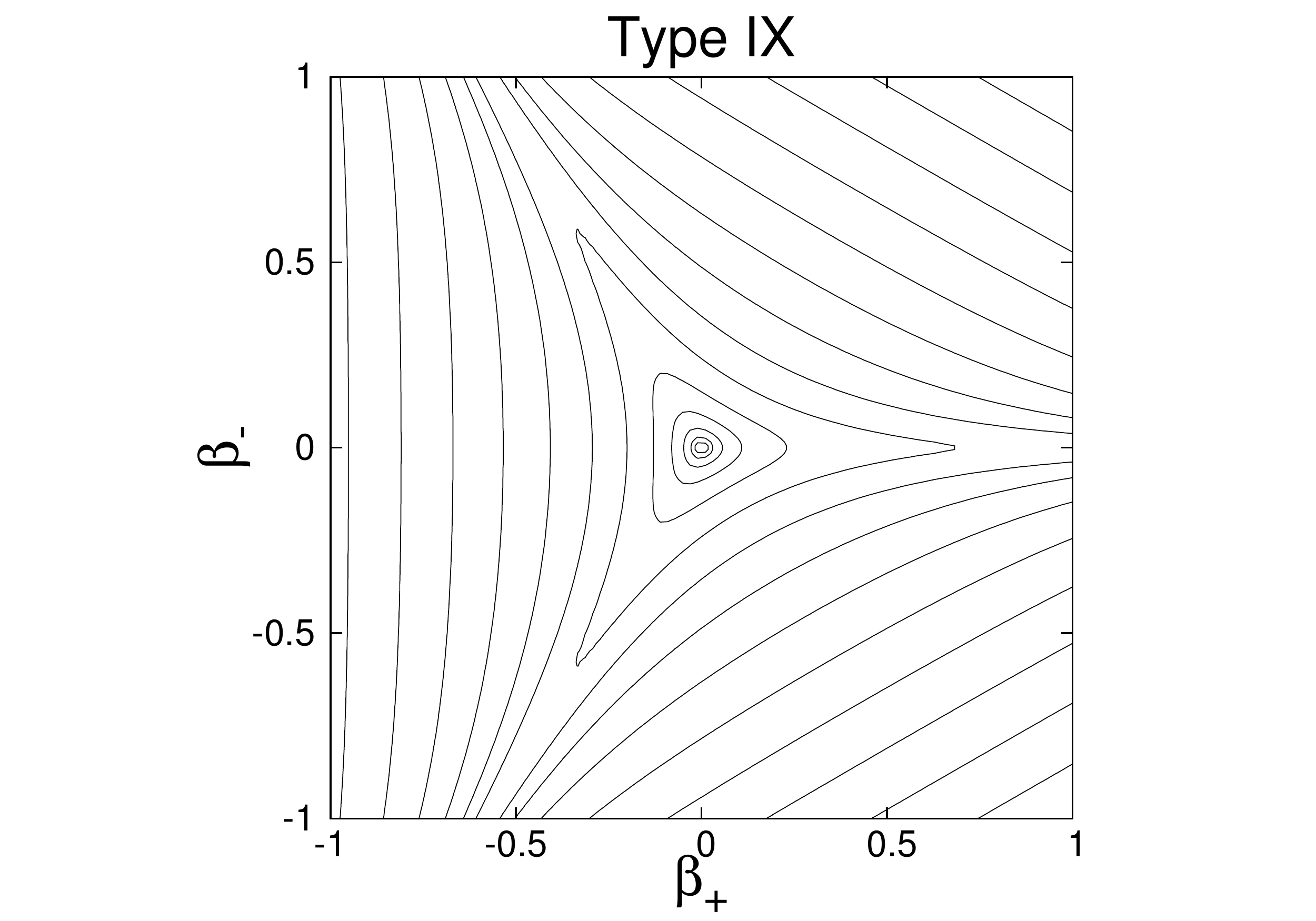}
  \end{center}
\end{minipage}
\end{tabular}
\caption{Equipotential curves for 
class
 A curvature potentials.}
\label{fig: equipot}
\end{figure}

It is often convenient to use the variables $\alpha(t), \beta_\pm(t)$, which are defined by:
\begin{eqnarray}
A=e^{\alpha+\beta_{+}+\sqrt{3}\beta_{-}}, \quad B=e^{\alpha+\beta_{+}-\sqrt{3}\beta_{-}}, \quad C=e^{\alpha-2\beta_{+}}. \label{eq: MisnerParam}
\end{eqnarray}
In terms of these coordinates, the curvature potentials are as shown in Table \ref{table: CurvaturePotentials}.  Note that the dependence on $\alpha$
is always of the form $V_c=e^{\alpha} v_c(\beta_\pm)$.  The coordinates of (\ref{eq: MisnerParam}) are called the Misner parameterization
\cite{MisnerMixmaster}.  The physical meaning of these dynamical variables is clear: $e^{\alpha}$ is the universe scale factor, and the $\beta_{\pm}$
characterize spatial anisotropy.  In terms of the Misner parameterization, the class A Hamiltonians have the following form:
\begin{equation}
H=N\left(\frac{1}{24}e^{-3 \alpha} \left( - \pi_{\alpha}^{2}+\pi_{\beta +}^{2}+\pi_{\beta -}^{2} \right)+2\Lambda e^{3\alpha}+V_c+V_m \right). 
\label{eq: Hamiltonian-alpha-beta}
\end{equation}
The equipotential curves for $V_c$ in the $\beta_+ - \beta_-$-plane are depicted in Figure \ref{fig: equipot}.

%%%%%%%%%%%%%%%%%%%%%%%%%%%%%%%%%%%%%%%%%%%%%%%%%%%%%%%%%%%%%%%%
\begin{table}[htbp]
\caption{The curvature potentials.  These potentials are calculated using a diagonal metric in a group-invariant basis: $g_{\mu\nu} =$ diag$\left(-N^2(t), A^2(t), B^2(t), C^2(t)\right)$.  The $\alpha(t), \beta_\pm(t)$ are defined by $A=e^{\alpha + \beta_+ + \sqrt{3}\beta_-}, B = e^{\alpha + \beta_+ - \sqrt{3}\beta_-},$ and $C=e^{\alpha - 2 \beta_+}$.
(Note: for calculations using a general form of the metric, see \cite{PonsShepley}.)}
\begin{ruledtabular}
\begin{tabular}{rl}
\textbf{Type} & $V_c=e^{\alpha}v_c(\beta_\pm)$\\
\hline
\textbf{I} & $V_I = 0$\\
\textbf{II} & $V_{II} = \frac{A^3}{2BC}=\frac{1}{2}e^{\alpha}e^{4(\beta_{+}+\sqrt{3}\beta_{-})}$ \\
\textbf{VI}$_{\bm{0}}$ & $V_{VI} = \frac{1}{2ABC} \left(A^2+B^2\right)^2 = 2e^{\alpha} e^{4\beta_{+}} \cosh^{2}(2\sqrt{3}\beta_{-})$ \\
\textbf{VII}$_{\bm{0}}$ & $V_{VII} = \frac{1}{2ABC} \left(A^2-B^2\right)^2 = 2e^{\alpha} e^{4\beta_{+}} \sinh^{2}(2\sqrt{3}\beta_{-})$ \\
\textbf{VIII} & $V_{VIII} = \frac{1}{2} \left(\frac{A^3}{BC}+\frac{B^3}{CA}+\frac{C^3}{AB}\right) - \left(\frac{AB}{C}-\frac{BC}{A}-\frac{CA}{B}\right)$ \\
 & $\qquad = e^{\alpha} \left(2e^{-2 \beta_{+}} \cosh(2\sqrt{3}\beta_{-}) - e^{4\beta_{+}} + e^{4\beta_{+}} \cosh(4\sqrt{3} \beta_{-}) + \frac{1}{2}e^{-8\beta_{+}}\right)$ \\
\textbf{IX} & $V_{IX} = \frac{1}{2} \left(\frac{A^3}{BC} + \frac{B^3}{CA} + \frac{C^3}{AB}\right) - \left(\frac{AB}{C} + \frac{BC}{A} + \frac{CA}{B}\right)$ \\
&\qquad $=e^{\alpha} \left(-2e^{-2 \beta_{+}} \cosh(2\sqrt{3}\beta_{-}) - e^{4\beta_{+}} + e^{4\beta_{+}} \cosh(4\sqrt{3} \beta_{-}) + \frac{1}{2}e^{-8\beta_{+}}\right)$
\end{tabular}
\end{ruledtabular}
\label{table: CurvaturePotentials}
\end{table}
%%%%%%%%%%%%%%%%%%%%%%%%%%%%%%%%%%%%%%%%%%%%%%%%%%%%%%%%%%%%%%%%

%%%%%%%%%%%%%%%%%%%%%%%%%%%%%%%%%%%%%%%%%%%%%%%%%%%%%%%%%%%%%%%%
%%%%%%%%%%%%%%%%%%%%%%%%%%%%%%%%%%%%%%%%%%%%%%%%%%%%%%%%%%%%%%%%

The term that comes from the presence of Vlasov matter is also a potential, which is denoted by $V_m$ in equation (\ref{eq: Hamiltonian-alpha-beta}).
The matter potential is 
\begin{equation}
V_m=2e^{-\alpha} \int d^{3}h\, f \sqrt{m^{2}e^{2\alpha}+\left(h_{1}^{2}\, e^{-2\beta_{+}-2\sqrt{3}\beta_{-}}+h_{2}^{2}\, e^{-2\beta_{+}+2\sqrt{3}\beta_{-}}+h_{3}^{2}\, e^{4\beta_{+}}\right)}\,. \label{eq: VlasovMatterPotential_MisnerParam}
\end{equation}
In massless cases, the dependence on $\alpha$ can be separated from that on $\beta_\pm$ since
\begin{equation}
V_m=e^{-\alpha}v_m(\beta_\pm)
\end{equation}
where
\begin{equation}
v_m=2 \int d^{3}h \, f\sqrt{h_{1}^{2}\, e^{-2\beta_{+}-2\sqrt{3}\beta_{-}}+h_{2}^{2}\, e^{-2\beta_{+}+2\sqrt{3}\beta_{-}}+h_{3}^{2}\, e^{4\beta_{+}}}\,.
\end{equation}

%%%%%%%%%%%%%%%
%%%%%%%%%%%%%%%

\section{Results: Vlasov effects}

\label{sec: results}

In this section, we give a description of how a cosmological model containing Vlasov matter may mimic or may differ from an apparently similar vacuum
model. The results will be described, but the details of the calculations --- both analytical and numerical --- will be left for a subsequent paper. 
The general behavior in the sense of qualitative cosmology will be described, where the effects of the potentials are either negligible or act as hard
walls.  Keep in mind that these results were obtained for the case where $\Lambda=0$ and $m=0$.

From the Hamiltonian (\ref{eq: Hamiltonian-alpha-beta}), we see that
\begin{eqnarray}
\dot{\alpha} &=& \hbbn{H}{\pi_\alpha} = -\frac{N}{12} e^{-3\alpha} \pi_\alpha\,,
\\
\dot{\beta}_\pm &=& \hbbn{H}{\pi_{\beta_\pm}} = \frac{N}{12} e^{-3\alpha} \pi_{\beta_\pm}\,,
\\
\dot{\pi}_\alpha &=& -\hbbn{H}{\alpha} = N \left[\frac{1}{8} e^{-3\alpha}\left( -\pi_\alpha^2 + \pi_{\beta_+}^2 + \pi_{\beta_-}^2 \right) - \hbbn{}{\alpha}(V_m + V_c)\right]\,.
\end{eqnarray}
Presuming that $\dot{\alpha}$ is monotonic, $\alpha$ is used as the ``time" variable.  In that case the dynamics of the model will be governed by 
the behavior of $\beta_\pm(\alpha)$.  The motion of the universe point (a point in the $\beta_+ - \beta_-$-plane) is with speed $w$,
using $\alpha$ as the ``time variable":
\begin{equation}
w^2 = \left( \frac{d\beta_+}{d\alpha} \right)^2 + \left( \frac{d\beta_-}{d\alpha} \right)^2.
\end{equation}
The Hamiltonian of equation (\ref{eq: Hamiltonian-alpha-beta}) is used in the Hamiltonian constraint (\ref{eq: HamiltonianABC}) to find
\begin{equation}
w^2 = 1 - 24e^{3\alpha}(V_c+V_m)/\pi_\alpha^2\,.
\label{eq: potential-wall-speed}
\end{equation}

When the potentials may be neglected, $w = 1$; this motion corresponds to the Kasner model.  In this sense the models may be thought of as a series of
Kasner epochs which transition from one Kasner state to another when the universe point interacts with a potential wall.  Note also that when the
potentials may be neglected, $\dot{\pi}_\alpha$ is proportional to $H$ and therefore vanishes.

The vacuum Bianchi models of Types II, VI$_0$, and IX will be compared with Type I models containing Vlasov matter. (See also
\cite{HeinzleUgglaTypeI}.) Since the Type I models do not have a curvature potential, this procedure emphasizes the effects of the matter potential. 
From Figure \ref{fig: equipot}, we see that the qualitative effects of the curvature potentials in the vacuum models II, VI$_0$, and IX are 
characterized by one wall, two walls, and three walls, respectively.  The equipotentials of Figure \ref{fig: equipot} define these walls, which move in 
time, and this movement will be described in more detail later.

For the matter potential, the distribution function $f$ (which in the Type I case is independent of time) is chosen to be that of cold,
counter-streaming matter:
\begin{equation}
f(h_i) = \frac{K}{8}\Bigl[\delta(h_1-a)+\delta(h_1+a)\Bigr] \Bigl[\delta(h_2-b)+\delta(h_2+b)\Bigr] \Bigl[\delta(h_3-c)+\delta(h_3+c)\Bigr]\,.
\label{2streamf}
\end{equation}
An example of the potential functions $\mu$, $\nu$ of equation (\ref{mu-nu-distribution}) may be easily calculated in this case, since in Type I, the $h_i $
are canonical variables:
\begin{eqnarray}
\mu &=& x^1\,,
\\
\nu &=&  \frac{K}{8}\Bigl[H(h_1-a)+H(h_1+a)\Bigr] \Bigl[\delta(h_2-b)+\delta(h_2+b)\Bigr] \Bigl[\delta(h_3-c)+\delta(h_3+c)\Bigr]\,,
\label{potfuncstypeI}
\end{eqnarray}
where $H$ is the Heaviside or step function.
Of course, other forms for $\mu$, $\nu$ which yield $f(h_i)$ are possible, and the gauge group of these 
possibilities will be presented in a subsequent paper.

$V_m$ for this model is
\begin{equation}
V_m=2Ke^{-\alpha}\sqrt{m^2e^{2\alpha}+a^{2}e^{-2\beta_{+}-2\sqrt{3}\beta_{-}}+b^{2}e^{-2\beta_{+}+2\sqrt{3}\beta_{-}}+c^{2}e^{4\beta_{+}}}
\,.
\end{equation}
This distribution function is for eight streams of particles, counter-streaming by pairs. The choice  (\ref{2streamf}) for $f$  is motivated by  the
cold, two-stream instability of plasma physics,  where a large amount of free energy is stored in an equilibrium state that is released by linear
instability.  Studies of counter-streaming matter in the Newtonian gravitational case show that similar instabilities exist, and even explosive
instabilities may be expected \cite{CastiMorrisonSpiegel,KM95}.   Although this particular choice  is still simplistic, it is consistent with the
Bianchi symmetry and,  importantly,  it  provides a tractable example of the kind of effects to be expected when matter is nonthermal and anisotropic
in momentum space. We note that although individual beams are cold, this model does have a nonzero stress-energy tensor, which, because this 
distribution function is even,  is diagonal. 

\begin{figure}[htbp]
\begin{tabular}{ccc}
\hspace{-1.0cm}
\begin{minipage}{0.3\hsize}
  \begin{center}
    \includegraphics[width=1.0\hsize]{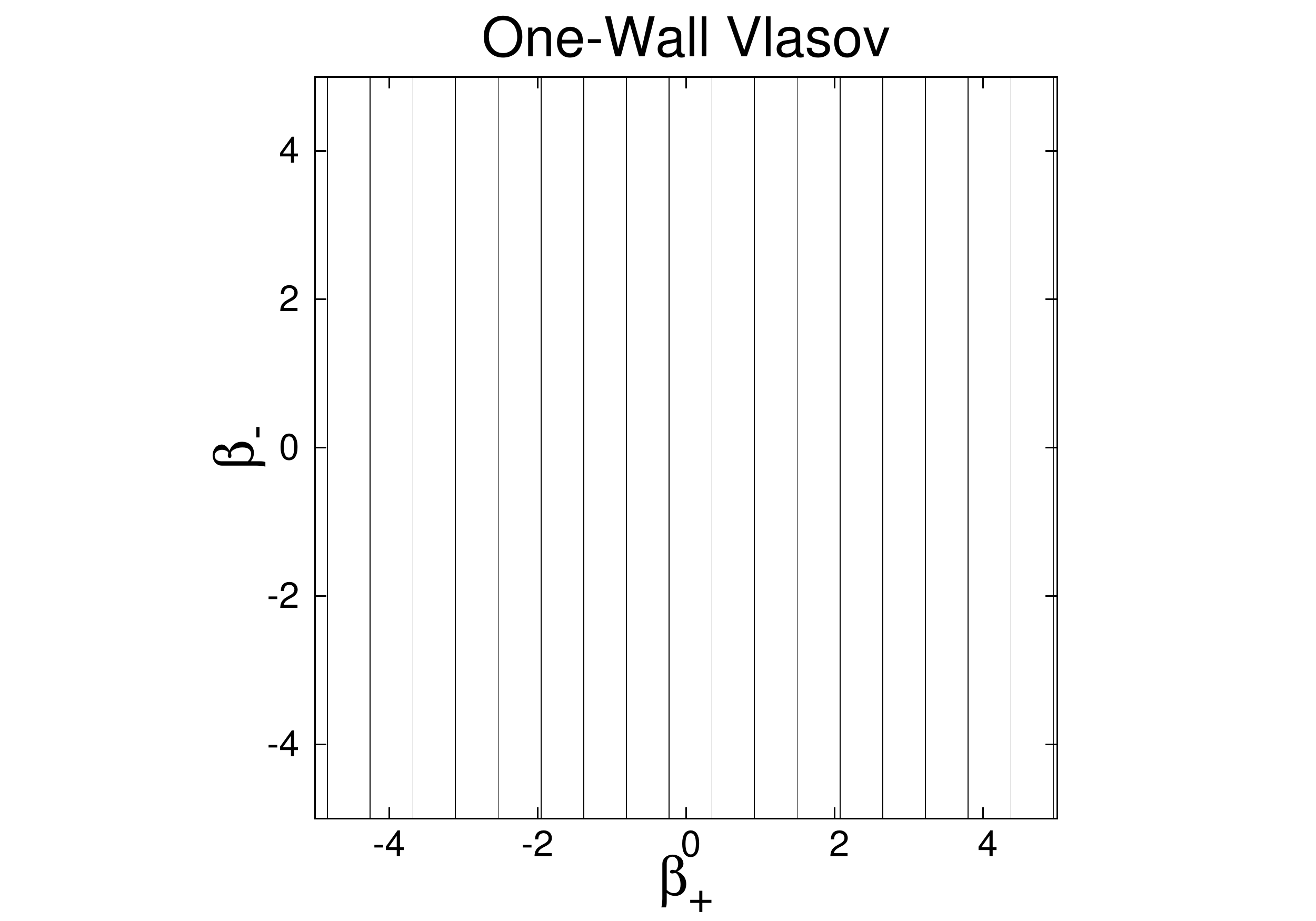}
  \end{center}
\end{minipage}
\hspace{-1cm}
\begin{minipage}{0.3\hsize}
  \begin{center}
    \includegraphics[width=1.0\hsize]{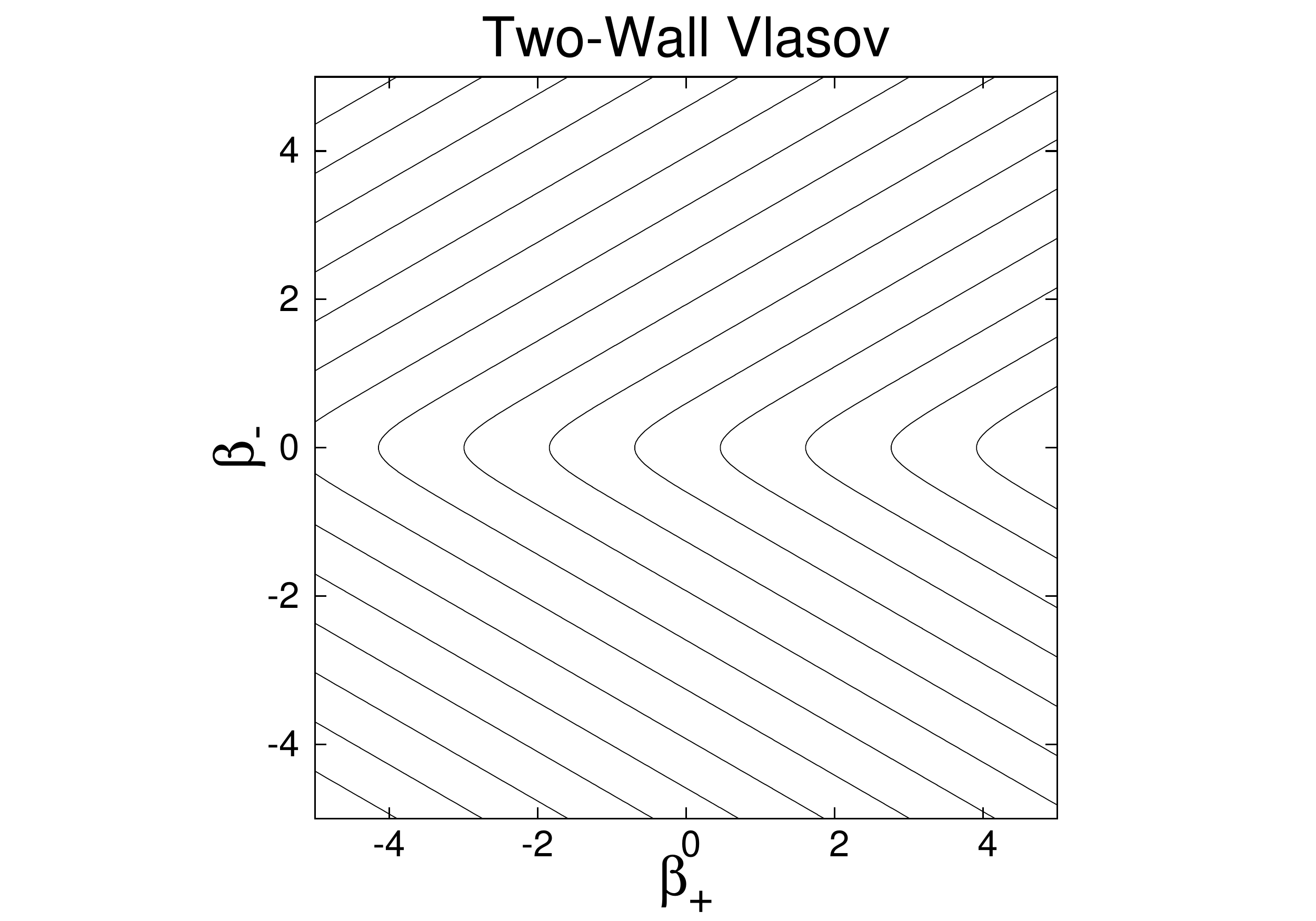}
  \end{center}
\end{minipage}
\hspace{-1cm}
\begin{minipage}{0.3\hsize}
  \begin{center}
    \includegraphics[width=1.0\hsize]{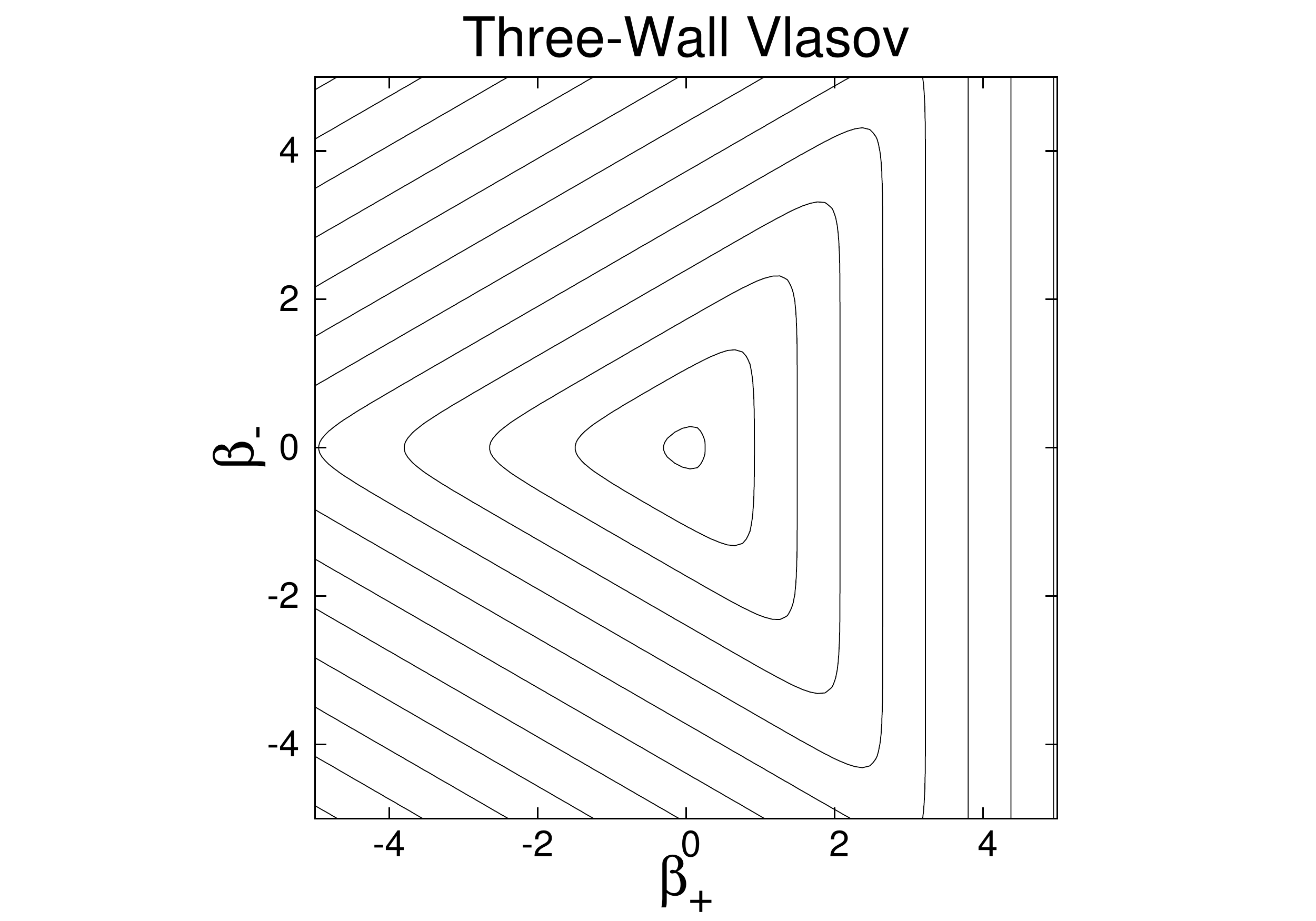}
  \end{center}
\end{minipage}
\end{tabular}
\caption{Equipotential curves for the Vlasov-matter potential with cold counter-streaming matter.}
\label{fig: equipot-vm}
\end{figure}

If $a = b = 0$, $c > 0$, the model is called one-wall Vlasov.  If $a = b > 0$, $c = 0$, it is called two-wall Vlasov.  Finally, if $a = b = c > 0$, it
is called three-wall Vlasov.  These are the only models considered for inclusion in this paper, with $m=0$ just for the sake of illustration.  The
forms of the matter potentials, labeled $V_1$, $V_2$, $V_3$, are illustrated in Figure \ref{fig: equipot-vm}.

%%%%%%%%%%%%%%%%%%%%%%%%%%%%%%%%%%%%%%%%%%%%%%%%%%%%%%%%%%%%%%%%
%%%%%%%%%%%%%%%%%%%%%%%%%%%%%%%%%%%%%%%%%%%%%%%%%%%%%%%%%%%%%%%%

These potentials are given analytically by the following:
\smallskip

\noindent One-wall Vlasov: $a=b=0, c>0$ with $m=0$
\begin{equation}
	V_1= 2Ke^{-\alpha}ce^{2\beta_+}.
\end{equation}
Two-wall Vlasov: $a=b \neq 0, c=0$ with $m=0$
\begin{equation}
	V_2= 2\sqrt{2}Kae^{-\alpha}e^{-\beta_+}\sqrt{\cosh(2\sqrt{3}\beta_-)}\,. 
\label{eq: potential two-wall Vlasov}
\end{equation}
Three-wall Vlasov: $a=b=c\neq 0$ with $m=0$
\begin{equation}
	V_3=2Kae^{-\alpha}\sqrt{2e^{-2\beta_+}\cosh(2\sqrt{3}\beta_-)+e^{4\beta_+}}\,.
\label{eq: potential three-wall Vlasov}
\end{equation}

These models of Vlasov matter in a Type I cosmology will be compared, respectively, to vacuum Bianchi models of Types II, VI$_0$, and IX (compare
Figures \ref{fig: equipot} and \ref{fig: equipot-vm}).  The curvature potentials for the vacuum models are labeled $V_{II}$, $V_{VI}$, and $V_{IX}$;
see Table \ref{table: CurvaturePotentials}.  The curvature potential for a Type I cosmology vanishes. Therefore in all cases when the universe point
has an energy substantially above the value of the potential, the model acts like a Kasner model.  The similarities and differences in the three
situations  described depend on whether  the universe point is strongly affected by the potential, in which case the universe point is said to
bounce off a potential wall or be influenced by it. 

The first pair is vacuum Type II compared to one-wall Vlasov (in Type I).  The one-wall effective potential $V_{\mathrm{eff}}$, from equation
(\ref{eq: potential-wall-speed}), is proportional to $e^{2\alpha+2\beta_+}$, so that the equipotential walls move with speed 
\begin{equation}
	w_{1} = 1.
\end{equation}
In contrast, the effective potential in the vacuum Type II case, from Table \ref{table: CurvaturePotentials}, is proportional to
$e^{4\alpha+4\left(\beta_+ + \sqrt{3}\beta_-\right)}$, so that the speed of the wall is
\begin{equation}
	w_{II} = \frac{1}{2}.
\end{equation}
Consequently the universe point in the vacuum Type II case will (in many cases) strike a potential wall and bounce (once) in a very well-defined
manner and then move away from the wall.  Qualitative cosmology is useful in this case.  In contrast, the universe point in the one-wall Vlasov model
and a potential wall both have the same speeds.  If the universe point is not moving exactly in the opposite direction of a wall, it will bounce. 
After the bounce, the universe point cannot escape the influence of the potential, since equation (\ref{eq: potential-wall-speed}) shows that its
velocity will in general be less than 1. Therefore qualitative cosmology is not a good approximation.  (Note,  the one-wall case corresponds to an invariant boundary in  the analysis of \cite{HeinzleUgglaTypeI}.)  

The two-wall Vlasov model and the Type VI$_0$ model also have a qualitatively similar potential wall structure.  The effective potentials for these
two cases, from equation (\ref{eq: potential-wall-speed}), is $e^{3\alpha}$ times the potentials listed in Table \ref{table: CurvaturePotentials} or
equation (\ref{eq: potential two-wall Vlasov}):
\begin{equation}
	V_{VI}^{eff} = 2e^{4\alpha}e^{4\beta_{+}}\cosh^2\left(2\sqrt{3}\beta_{-}\right)
\end{equation}
\begin{equation}
	V_{2}^{eff} = ke^{2\alpha}e^{-\beta_{+}}\sqrt{\cosh\left(2\sqrt{3}\beta_{-}\right)}\,.
\end{equation}	 
The speeds of the vertices of the equipotentials (that is, when $\beta_{-}=0$) are:
\begin{equation}
	w_{VI}^{vert} = 1
\end{equation}
\begin{equation}
	w_{2}^{vert} = 2.
\end{equation}
The speeds of the walls, however, are governed by their slopes far from the vertices.  At large $\beta_{-}$ the effective potentials are approximated by
\begin{equation}
	V_{VI}^{eff} \approx e^{4\alpha}e^{4\beta_{+}+4\sqrt{3}\beta_{-}}
\end{equation}
\begin{equation}
	V_{2}^{eff} \approx \frac{k}{\sqrt{2}}\, e^{2\alpha}e^{-\beta_{+}+\sqrt{3}\beta_{-}}\,.
\end{equation}
The speeds of the walls, therefore are
\begin{equation}
	w_{VI} = \frac{1}{2}
\end{equation}
\begin{equation}
	w_{2} = 1
\end{equation}
(as in the previous case).  Therefore the same comments about the universe point hitting the potential walls apply: Qualitative cosmology is a good
approximation for the Type VI$_0$ case but not for the two-wall Vlasov case. (Note,  as with the one wall case, the two-wall case  corresponds to an invariant boundary  in  the analysis of \cite{HeinzleUgglaTypeI}.)

The three-wall Vlasov and the Type IX vacuum Bianchi models are somewhat more interesting.  The Type IX model eventually collapses.  Note also that
the potential walls have channels (see Figure \ref{fig: equipot}),  which together with the steepness of the walls produce what Misner calls the Mixmaster Model \cite{MisnerMixmaster}. The
three-wall Vlasov model has an effective potential of $e^{3\alpha}$ times the potential of equation (\ref{eq: potential three-wall Vlasov}):
\begin{equation}
	V_{3}^{eff} = ke^{2\alpha} \sqrt{2e^{-2\beta_{+}}\cosh\left(2\sqrt{3}\beta_{-}\right) + e^{4\beta_{+}}}.
\end{equation}
The speed of each vertex (for example when $\beta_{-}=0$, along the negative $\beta_{+}$-axis) is 
\begin{equation}
	w_{3}^{vert} = 2.
\end{equation}
The speed of each wall is found by looking at the vertical wall when $\beta_{-}=0$, along the positive $\beta_{+}$-axis:
\begin{equation}
	w_{3} = 1
\end{equation}
Again it is seen that the details of the motion of the universe point cannot be approximated by qualitative cosmology.

The details of the motions in the three Vlasov models as contrasted with the corresponding vacuum models,  in spite of the apparent similarity of the
potentials, will be left to a subsequent paper. However, in the two-wall case the speed of the vertices of the equipotentials shows that 
partial isotropization takes place (namely $\beta_{-}\rightarrow0$).  The three-wall Vlasov model has complete 
isotropization: The universe point goes to the origin.

To examine this isotropization more closely, we first look at the stress-energy tensor, equation (\ref{eq: Einstein}), for the three-wall case in a
Type I model.  In this case the $p_i$ are the same as the $h_i$, and in our gauge, $\sqrt{-g}=NABC$.  Moreover, the on-shell condition implies
\begin{equation}
	\bar{p}^0=\frac{1}{N}\sqrt{m^2 + (p_1/A)^2+ (p_2/B)^2+ (p_3/C)^2}.
\end{equation}
The stress-energy tensor is diagonal, with components:
\begin{equation}
	T_{00} = N^2 \frac{K}{ABC}\Pi
\end{equation}
\begin{equation}
	T_{11} = A^2 \frac{K}{ABC}\frac{a^2}{\Pi A^2}
\end{equation}
\begin{equation}
	T_{22} = B^2 \frac{K}{ABC}\frac{b^2}{\Pi B^2}
\end{equation}
\begin{equation}
	T_{33} = C^2 \frac{K}{ABC}\frac{c^2}{\Pi C^2}\,,
\end{equation}
where the constants $K,a,b,c$ are defined in equation (\ref{2streamf}) and where
\begin{equation}
	\Pi: = \sqrt{m^2 +(a/A)^2 +(b/B)^2 +(c/C)^2}.
\end{equation}
In the limiting case of $A=B=C$ (when $\beta_\pm=0$), and if it is approximately true that $a=b=c$, then the stress-energy tensor becomes that of a
perfect fluid with $\W$ the energy density and $\P$ the pressure, given by
\begin{equation}
	\W= \frac{K\Pi}{A^3}, \hspace{10 pt} \P=\frac{Ka^2}{A^5 \Pi}.
\end{equation}
Their ratio is the equation of state:
\begin{equation}
	\frac{\P}{\W} = \frac{1}{3+(mA/a)^2}.
\end{equation}
If $m\neq 0$, the pressure goes to zero at large $A$ (so-called ``dust"), and if $m=0$, the factor $1/3$ is that of a fluid of massless particles.  This is consistent with  results of  \cite{Rendall,HeinzleUgglaTypeI} that indicate  that models of a class, which includes our three-wall model,   will isotropize in the asymptotic future.

As for the action integral, the $h$ integration can be done to get, in our diagonal Type I model, 
\begin{eqnarray}
	S_{{\mathrm{matter}}}=-2K\int N\Pi\,dt=-2K\int N\sqrt{m^2 + (a/A)^2 + (b/B)^2 + (c/C)^2}\, dt  \nonumber\\
=-2K\int \frac{1}{\sqrt{-g^{00}}}\sqrt{m^2 + a^2 g^{11} + b^2 g^{22} + c^2 g^{33} }\,dt.
\end{eqnarray}
The variations with respect to $g^{\mu\nu}$ give the stress-energy tensor components displayed before.

%%%%%%%%%%%%%%%
%%%%%%%%%%%%%%%

\section{Conclusion}

In this work, the matter action for the Vlasov-Einstein system for equal mass particles is constructed. As an illustration of this action, we found a
new kind of potential $V_m$ for Vlasov matter in the anisotropy plane of spatially-homogeneous models (those of Bianchi class A with diagonal metric).

Then, in order to investigate the similarities and the differences between the Vlasov-matter potential and the curvature potential, the Type I
universe with cold, counter-streaming matter and the corresponding vacuum Bianchi models were compared on the basis of the shape of the potentials.    Both kinds of  potentials, the curvature potentials $V_c$ which arise from the requirement of spatial homogeneity in a
vacuum model and the $V_m$, were classified by the number of approximately straight equipotential sides.  The $V_c$ in models of Type II has one; in
Types VI$_0$ and VII$_0$ the $V_c$ have two; and in Type VIII and IX the $V_c$ have three. Their counterparts, one-wall, two-wall, and three-wall
Vlasov-matter potentials were constructed by selecting a cold, counter-streaming matter distribution function in a Type I model.  The $V_c$ of a
vacuum Type II model was compared with the one-wall Vlasov model.  The $V_c$ of a vacuum Type VI$_0$ model was compared with the two-wall Vlasov
model.  The $V_c$ of a vacuum Type IX model was compared with the three-wall Vlasov model, since they share triangular symmetry, in spite of the 
absence of channels in the Vlasov case.

Aside from questions of recollapse and direction of  increase or decrease in anisotropy, it was seen that the $V_c$ and $V_m$ had different
effects because of the relative speeds of the walls and the universe point described by $\beta_\pm(\alpha)$.  The dynamics of the vacuum Bianchi
cosmologies can be approximated as a series of Kasner eras where the point representing the metric in the $\beta_\pm$ space moves freely except for
bounces off the potential walls.  Because $V_m$ is not as sharp as $V_c$, and because the speed of its walls was comparable to the speed of the
universe point, it would be difficult to separate out a potential-free region:  Consequently, contact with $V_m$ could not be characterized as a
transition from a Kasner state to another Kasner state.  In a subsequent paper, details of the effects of Vlasov matter on the evolution of anisotropy
will be described both for our present case of cold,  counter-streaming matter  and matter which may be considered warm.

Early work on this subject was done by Misner \cite{MisnerViscosity} and Matzner \cite{Matzner}.  Their idea was to ask whether a kinetic 
theory of matter would cause anisotropy to decay faster than it would in a model with perfect fluid.  They took a general approach, in which 
the distribution function had a thermal structure because their massless particles (which they called neutrinos) were in thermal equilibrium 
with other matter until a particular time when the temperature dropped below a critical value.  Then there was a period of transition followed 
by the present epoch, when the neutrinos are essentially collisionless.   Misner looked at Type I models.  Matzner looked at Type IX and 
Type V models, with and without rotation.  They were able to get a general idea of a potential due to the particles. This potential was a 
triangular one, possibly with fewer than three walls, depending on the form of the distribution function.  Our methodology differs from theirs.
Our distribution function in our main example also differs, but our general results are in accord with theirs.

Rendall \cite{Rendall} has also studied the effects of Vlasov matter in a Bianchi Type I model.  His results cover a more general set of 
distribution functions than we considered in our particular example.  In so far as our results can be compared, they agree.  Rendall did 
not use a Hamiltonian formalism, and we feel that subsequent studies will benefit from our approach.

In conclusion, we list several possible areas for future work. Examples of minor projects are as follows: First of all, the distribution functions of
Ehlers type were used in this work, and the dynamics of the more general distribution functions given in equation (3.25) of Chapter 3 of
\cite{TakahideDissertation} should be treated. It would be of interest to perform numerical computations for these more general $f(t, h_k)$. Second,
our system was simplified by assuming a diagonal metric (\ref{eq: DiagonalMetric}), and it could be generalized to include off-diagonal elements and
possibly even rotating models. Third, our examples focused on massless cases for simplicity, and the massive case should be investigated.

There are at least two major projects. First, we propose to study further the problem of a Bianchi Type IX model with non-zero cosmological constant
and Vlasov matter. This case has Einstein's static universe as an unstable fixed point. Whether this point is a chaotic scatterer or not for dust
filled models was discussed in
\cite{TypeIXChaoticScatterer1,TypeIXChaoticScatterer2,TypeIXChaoticScatterer3,TypeIXChaoticScatterer4,TypeIXChaoticScatterer5}, which has no
additional anisotropic potential.  The inclusion of $V_m$ for Vlasov matter might change the behavior of the scatterer, and introduce possible chaotic
effects.  Second, we propose the problem of the Type I universe with the Vlasov-matter potential $V_m$ in the presence of a negative cosmological
constant. In this case, the universe eventually re-collapses and the potential walls recede, just as for the vacuum Type IX case. However, unlike what
happens in the vacuum Type IX case, the universe point cannot catch up with the retreating potential walls of $V_m$, since they can move with a speed
comparable to that of the universe point. Therefore, with a negative cosmological constant, the system can be characterized as a transition from an
initial Kasner state to a final Kasner state with the intermediate dynamics being complicated: that is, the system can be viewed as a scattering
problem.

\nocite{*}
\bibliographystyle{apsrev}      	% Here the bibliography
%\bibliographystyle{plain}
%\bibliography{V_Epaper}        			% is inserted.
\bibliography{V_E-paper-v9}

\begin{thebibliography}{41}
\expandafter\ifx\csname natexlab\endcsname\relax\def\natexlab#1{#1}\fi
\expandafter\ifx\csname bibnamefont\endcsname\relax
  \def\bibnamefont#1{#1}\fi
\expandafter\ifx\csname bibfnamefont\endcsname\relax
  \def\bibfnamefont#1{#1}\fi
\expandafter\ifx\csname citenamefont\endcsname\relax
  \def\citenamefont#1{#1}\fi
\expandafter\ifx\csname url\endcsname\relax
  \def\url#1{\texttt{#1}}\fi
\expandafter\ifx\csname urlprefix\endcsname\relax\def\urlprefix{URL }\fi
\providecommand{\bibinfo}[2]{#2}
\providecommand{\eprint}[2][]{\url{#2}}

\bibitem[{\citenamefont{{Carroll}}(1997)}]{SeanCarroll}
\bibinfo{author}{\bibfnamefont{S.~M.} \bibnamefont{{Carroll}}}
  (\bibinfo{year}{1997}), \bibinfo{note}{online lecture notes},
  \eprint{arXiv:gr-qc/9712019v1}.

\bibitem[{\citenamefont{{Carroll}}(2004)}]{SeanCarroll2}
\bibinfo{author}{\bibfnamefont{S.~M.} \bibnamefont{{Carroll}}},
  \emph{\bibinfo{title}{Spacetime and geometry : an introduction to general
  relativity}} (\bibinfo{publisher}{Addison Wesley}, \bibinfo{year}{2004}).

\bibitem[{\citenamefont{{Wald}}(1984)}]{Wald}
\bibinfo{author}{\bibfnamefont{R.~M.} \bibnamefont{{Wald}}},
  \emph{\bibinfo{title}{{General Relativity}}} (\bibinfo{publisher}{The
  University of Chicago Press}, \bibinfo{year}{1984}).

\bibitem[{\citenamefont{{Ryan} and {Shepley}}(1975)}]{RyanShepley}
\bibinfo{author}{\bibfnamefont{M.~P.} \bibnamefont{{Ryan}}} \bibnamefont{and}
  \bibinfo{author}{\bibfnamefont{L.~C.} \bibnamefont{{Shepley}}},
  \emph{\bibinfo{title}{{Homogeneous Relativistic Cosmologies}}}
  (\bibinfo{publisher}{Princeton University Press}, \bibinfo{year}{1975}).

\bibitem[{\citenamefont{Stewart}(1971)}]{Stewart}
\bibinfo{author}{\bibfnamefont{J.~M.} \bibnamefont{Stewart}},
  \emph{\bibinfo{title}{Non-equilibrium relativistic kinetic theory}},
  vol.~\bibinfo{volume}{10} of \emph{\bibinfo{series}{Lecture Notes in
  Physics}} (\bibinfo{publisher}{Springer-Verlag}, \bibinfo{year}{1971}).

\bibitem[{\citenamefont{{Ehlers}}(1971)}]{EhlersKineticTheory}
\bibinfo{author}{\bibfnamefont{J.}~\bibnamefont{{Ehlers}}}, in
  \emph{\bibinfo{booktitle}{General Relativity and Cosmology}}, edited by
  \bibinfo{editor}{\bibfnamefont{R.~K.} \bibnamefont{Sachs}}
  (\bibinfo{publisher}{Academic Press}, \bibinfo{year}{1971}), pp.
  \bibinfo{pages}{1--70}.

\bibitem[{\citenamefont{{Berezdivin} and {Sachs}}(1970)}]{BerezSachs}
\bibinfo{author}{\bibfnamefont{R.}~\bibnamefont{{Berezdivin}}}
  \bibnamefont{and} \bibinfo{author}{\bibfnamefont{R.~K.}
  \bibnamefont{{Sachs}}}, in \emph{\bibinfo{booktitle}{Relativity}}, edited by
  \bibinfo{editor}{\bibfnamefont{M.}~\bibnamefont{{Carmeli}}},
  \bibinfo{editor}{\bibfnamefont{S.~I.} \bibnamefont{{Fickler}}},
  \bibnamefont{and} \bibinfo{editor}{\bibfnamefont{L.}~\bibnamefont{{Witten}}}
  (\bibinfo{publisher}{Plenum Press}, \bibinfo{year}{1970}), pp.
  \bibinfo{pages}{125--131}.

\bibitem[{\citenamefont{{Taub}}(1951)}]{Taub}
\bibinfo{author}{\bibfnamefont{A.~H.} \bibnamefont{{Taub}}},
  \bibinfo{journal}{Ann. Math.} \textbf{\bibinfo{volume}{53}},
  \bibinfo{pages}{472} (\bibinfo{year}{1951}).

\bibitem[{\citenamefont{Dirac}(1958)}]{Dirac}
\bibinfo{author}{\bibfnamefont{P.}~\bibnamefont{Dirac}},
  \bibinfo{journal}{Proc. Roy. Soc. A} \textbf{\bibinfo{volume}{246}},
  \bibinfo{pages}{333} (\bibinfo{year}{1958}).

\bibitem[{\citenamefont{Arnowitt et~al.}(1962)\citenamefont{Arnowitt, Deser,
  and Misner}}]{ADM}
\bibinfo{author}{\bibfnamefont{R.}~\bibnamefont{Arnowitt}},
  \bibinfo{author}{\bibfnamefont{S.}~\bibnamefont{Deser}}, \bibnamefont{and}
  \bibinfo{author}{\bibfnamefont{C.}~\bibnamefont{Misner}}, in
  \emph{\bibinfo{booktitle}{Gravitation - An Introduction to Current
  Research}}, edited by
  \bibinfo{editor}{\bibfnamefont{L.}~\bibnamefont{Witten}}
  (\bibinfo{publisher}{John Wiley}, \bibinfo{year}{1962}), pp.
  \bibinfo{pages}{227--265}.

\bibitem[{\citenamefont{{Takahide Okabe}}(2008)}]{TakahideDissertation}
\bibinfo{author}{\bibnamefont{{Takahide Okabe}}}, Ph.D. thesis,
  \bibinfo{school}{The University of Texas at Austin} (\bibinfo{year}{2008}).

\bibitem[{\citenamefont{Andr\'{e}asson}(2005)}]{Andreasson}
\bibinfo{author}{\bibfnamefont{H.}~\bibnamefont{Andr\'{e}asson}}
  (\bibinfo{year}{2005}), \bibinfo{note}{{L}iving Reviews in Relativity},
  \eprint{http://relativity.livingreviews.org/Articles/lrr-2005-2/}.

\bibitem[{\citenamefont{{Ellis} and {MacCallum}}(1969)}]{MacCallum3}
\bibinfo{author}{\bibfnamefont{G.~F.~R.} \bibnamefont{{Ellis}}}
  \bibnamefont{and} \bibinfo{author}{\bibfnamefont{M.~A.~H.}
  \bibnamefont{{MacCallum}}}, \bibinfo{journal}{Comm. Math. Phys.}
  \textbf{\bibinfo{volume}{12}}, \bibinfo{pages}{108} (\bibinfo{year}{1969}).

\bibitem[{\citenamefont{Ryan}(1970)}]{RyanDissertation}
\bibinfo{author}{\bibfnamefont{M.~P.} \bibnamefont{Ryan}}, Ph.D. thesis,
  \bibinfo{school}{The University of Maryland} (\bibinfo{year}{1970}).

\bibitem[{\citenamefont{Collins and Stewart}(1971)}]{CollinsStewart}
\bibinfo{author}{\bibfnamefont{C.~B.} \bibnamefont{Collins}} \bibnamefont{and}
  \bibinfo{author}{\bibfnamefont{J.~M.} \bibnamefont{Stewart}},
  \bibinfo{journal}{Mon. Not. Roy. Astron. Soc.}
  \textbf{\bibinfo{volume}{153}}, \bibinfo{pages}{419} (\bibinfo{year}{1971}).

\bibitem[{\citenamefont{{Collins}}(1971)}]{Collins}
\bibinfo{author}{\bibfnamefont{C.~B.} \bibnamefont{{Collins}}},
  \bibinfo{journal}{Comm. Math. Phys.} \textbf{\bibinfo{volume}{23}},
  \bibinfo{pages}{137} (\bibinfo{year}{1971}).

\bibitem[{\citenamefont{{Morrison}}(1981)}]{morrison81}
\bibinfo{author}{\bibfnamefont{P.~J.} \bibnamefont{{Morrison}}},
  \bibinfo{journal}{Princeton Plasma Physics Laboratory Report}
  \textbf{\bibinfo{volume}{PPPL--1788}}, \bibinfo{pages}{1}
  (\bibinfo{year}{1981}).

\bibitem[{\citenamefont{Morrison}(1982)}]{morrison82}
\bibinfo{author}{\bibfnamefont{P.~J.} \bibnamefont{Morrison}}, in
  \emph{\bibinfo{booktitle}{Mathematical Methods in Hydrodynamics and
  Integrability in Dynamical Systems}}, edited by
  \bibinfo{editor}{\bibfnamefont{M.}~\bibnamefont{Tabor}} \bibnamefont{and}
  \bibinfo{editor}{\bibfnamefont{Y.}~\bibnamefont{Treve}}
  (\bibinfo{publisher}{American Institute of Physics}, \bibinfo{year}{1982}),
  no.~\bibinfo{number}{88} in \bibinfo{series}{AIP conference proceedings}, pp.
  \bibinfo{pages}{13--46}.

\bibitem[{\citenamefont{{Ye} and {Morrison}}(1992)}]{MorrisonYe}
\bibinfo{author}{\bibfnamefont{H.}~\bibnamefont{{Ye}}} \bibnamefont{and}
  \bibinfo{author}{\bibfnamefont{P.~J.} \bibnamefont{{Morrison}}},
  \bibinfo{journal}{Phys. Fluids B: Plasma Phys.} \textbf{\bibinfo{volume}{4}},
  \bibinfo{pages}{771} (\bibinfo{year}{1992}).

\bibitem[{\citenamefont{{Kandrup} and {O'Neill}}(1994)}]{KandrupNeill}
\bibinfo{author}{\bibfnamefont{H.~E.} \bibnamefont{{Kandrup}}}
  \bibnamefont{and}
  \bibinfo{author}{\bibfnamefont{E.}~\bibnamefont{{O'Neill}}},
  \bibinfo{journal}{Phys. Rev. D} \textbf{\bibinfo{volume}{49}},
  \bibinfo{pages}{5115} (\bibinfo{year}{1994}).

\bibitem[{\citenamefont{Morrison}(1980)}]{Morrison80}
\bibinfo{author}{\bibfnamefont{P.~J.} \bibnamefont{Morrison}},
  \bibinfo{journal}{Phys. Lett.} \textbf{\bibinfo{volume}{80A}},
  \bibinfo{pages}{383} (\bibinfo{year}{1980}).

\bibitem[{\citenamefont{{Kandrup} and {Morrison}}(1993)}]{KandrupMorrison}
\bibinfo{author}{\bibfnamefont{H.~E.} \bibnamefont{{Kandrup}}}
  \bibnamefont{and} \bibinfo{author}{\bibfnamefont{P.~J.}
  \bibnamefont{{Morrison}}}, \bibinfo{journal}{Ann. Phys. (NY)}
  \textbf{\bibinfo{volume}{225}}, \bibinfo{pages}{114} (\bibinfo{year}{1993}).

\bibitem[{\citenamefont{{Marsden} et~al.}(1986)\citenamefont{{Marsden},
  {Montgomery}, {Morrison}, and {Thompson}}}]{MMMT86}
\bibinfo{author}{\bibfnamefont{J.~E.} \bibnamefont{{Marsden}}},
  \bibinfo{author}{\bibfnamefont{R.}~\bibnamefont{{Montgomery}}},
  \bibinfo{author}{\bibfnamefont{P.~J.} \bibnamefont{{Morrison}}},
  \bibnamefont{and} \bibinfo{author}{\bibfnamefont{W.~B.}
  \bibnamefont{{Thompson}}}, \bibinfo{journal}{Ann. Phys. (NY)}
  \textbf{\bibinfo{volume}{169}}, \bibinfo{pages}{29} (\bibinfo{year}{1986}).

\bibitem[{\citenamefont{{Anderson} and {Spiegel}}(1972)}]{AndersonSpiegel}
\bibinfo{author}{\bibfnamefont{J.~L.} \bibnamefont{{Anderson}}}
  \bibnamefont{and} \bibinfo{author}{\bibfnamefont{E.~A.}
  \bibnamefont{{Spiegel}}}, \bibinfo{journal}{Astrophys. J.}
  \textbf{\bibinfo{volume}{171}}, \bibinfo{pages}{127} (\bibinfo{year}{1972}).

\bibitem[{\citenamefont{{Maartens} and {Maharaj}}(1985)}]{MaartensMaharaj1985}
\bibinfo{author}{\bibfnamefont{R.}~\bibnamefont{{Maartens}}} \bibnamefont{and}
  \bibinfo{author}{\bibfnamefont{S.~D.} \bibnamefont{{Maharaj}}},
  \bibinfo{journal}{J. Math. Phys.} \textbf{\bibinfo{volume}{26}},
  \bibinfo{pages}{2869} (\bibinfo{year}{1985}).

\bibitem[{\citenamefont{{Friedrichsen III}}(2000)}]{JamesDissertation}
\bibinfo{author}{\bibfnamefont{J.~E.} \bibnamefont{{Friedrichsen III}}}, Ph.D.
  thesis, \bibinfo{school}{The University of Texas at Austin}
  (\bibinfo{year}{2000}).

\bibitem[{\citenamefont{{MacCallum}}(1972)}]{MacCallum2}
\bibinfo{author}{\bibfnamefont{M.~A.~H.} \bibnamefont{{MacCallum}}},
  \bibinfo{journal}{Phys. Lett.} \textbf{\bibinfo{volume}{40A}},
  \bibinfo{pages}{385} (\bibinfo{year}{1972}).

\bibitem[{\citenamefont{{MacCallum} et~al.}(1970)\citenamefont{{MacCallum},
  {Stewart}, and G.}}]{MacCallum4}
\bibinfo{author}{\bibfnamefont{M.~A.~H.} \bibnamefont{{MacCallum}}},
  \bibinfo{author}{\bibfnamefont{J.~M.} \bibnamefont{{Stewart}}},
  \bibnamefont{and} \bibinfo{author}{\bibfnamefont{S.~B.} \bibnamefont{G.}},
  \bibinfo{journal}{Comm. Math. Phys.} \textbf{\bibinfo{volume}{17}},
  \bibinfo{pages}{343} (\bibinfo{year}{1970}).

\bibitem[{\citenamefont{{Misner}}(1969)}]{MisnerMixmaster}
\bibinfo{author}{\bibfnamefont{C.~W.} \bibnamefont{{Misner}}},
  \bibinfo{journal}{Phys. Rev. Lett.} \textbf{\bibinfo{volume}{22}},
  \bibinfo{pages}{1071} (\bibinfo{year}{1969}).

\bibitem[{\citenamefont{Pons and Shepley}(1998)}]{PonsShepley}
\bibinfo{author}{\bibfnamefont{J.~M.} \bibnamefont{Pons}} \bibnamefont{and}
  \bibinfo{author}{\bibfnamefont{L.~C.} \bibnamefont{Shepley}},
  \bibinfo{journal}{Phys. Rev. D} \textbf{\bibinfo{volume}{58}},
  \bibinfo{pages}{024001} (\bibinfo{year}{1998}).

\bibitem[{\citenamefont{{Heinzle} and {Uggla}}(2006)}]{HeinzleUgglaTypeI}
\bibinfo{author}{\bibfnamefont{J.~M.} \bibnamefont{{Heinzle}}}
  \bibnamefont{and} \bibinfo{author}{\bibfnamefont{C.}~\bibnamefont{{Uggla}}},
  \bibinfo{journal}{Class. Quant. Grav.} \textbf{\bibinfo{volume}{23}},
  \bibinfo{pages}{3463} (\bibinfo{year}{2006}).

\bibitem[{\citenamefont{{Casti} et~al.}(1998)\citenamefont{{Casti}, {Morrison},
  and {Spiegel}}}]{CastiMorrisonSpiegel}
\bibinfo{author}{\bibfnamefont{A.~R.~R.} \bibnamefont{{Casti}}},
  \bibinfo{author}{\bibfnamefont{P.~J.} \bibnamefont{{Morrison}}},
  \bibnamefont{and} \bibinfo{author}{\bibfnamefont{E.~A.}
  \bibnamefont{{Spiegel}}}, \bibinfo{journal}{Ann. NY Acad. Sci.}
  \textbf{\bibinfo{volume}{867}}, \bibinfo{pages}{93} (\bibinfo{year}{1998}).

\bibitem[{\citenamefont{Kueny and Morrison}(1995)}]{KM95}
\bibinfo{author}{\bibfnamefont{C.~S.} \bibnamefont{Kueny}} \bibnamefont{and}
  \bibinfo{author}{\bibfnamefont{P.~J.} \bibnamefont{Morrison}},
  \bibinfo{journal}{Phys. Plasmas} \textbf{\bibinfo{volume}{2}},
  \bibinfo{pages}{1926} (\bibinfo{year}{1995}).

\bibitem[{\citenamefont{Rendall}(1996)}]{Rendall}
\bibinfo{author}{\bibfnamefont{A.~D.} \bibnamefont{Rendall}},
  \bibinfo{journal}{J. Math. Phys.} \textbf{\bibinfo{volume}{37}},
  \bibinfo{pages}{438} (\bibinfo{year}{1996}).

\bibitem[{\citenamefont{Misner}(1971)}]{MisnerViscosity}
\bibinfo{author}{\bibfnamefont{C.~W.} \bibnamefont{Misner}},
  \bibinfo{journal}{Astrophys. J.} \textbf{\bibinfo{volume}{65}},
  \bibinfo{pages}{431} (\bibinfo{year}{1971}).

\bibitem[{\citenamefont{Matzner}(1971)}]{Matzner}
\bibinfo{author}{\bibfnamefont{R.~A.} \bibnamefont{Matzner}},
  \bibinfo{journal}{Ann. Phys. (NY)} \textbf{\bibinfo{volume}{153}},
  \bibinfo{pages}{438} (\bibinfo{year}{1971}).

\bibitem[{\citenamefont{{de Oliveira} et~al.}(1997)\citenamefont{{de Oliveira},
  {Dami{\~a}o Soares}, and {Stuchi}}}]{TypeIXChaoticScatterer1}
\bibinfo{author}{\bibfnamefont{H.~P.} \bibnamefont{{de Oliveira}}},
  \bibinfo{author}{\bibfnamefont{I.}~\bibnamefont{{Dami{\~a}o Soares}}},
  \bibnamefont{and} \bibinfo{author}{\bibfnamefont{T.~J.}
  \bibnamefont{{Stuchi}}}, \bibinfo{journal}{Phys. Rev. D}
  \textbf{\bibinfo{volume}{56}}, \bibinfo{pages}{730} (\bibinfo{year}{1997}).

\bibitem[{\citenamefont{{de Oliveira} et~al.}(2002)\citenamefont{{de Oliveira},
  {Ozorio de Almeida}, {Dami{\~a}o Soares}, and
  {Tonini}}}]{TypeIXChaoticScatterer2}
\bibinfo{author}{\bibfnamefont{H.~P.} \bibnamefont{{de Oliveira}}},
  \bibinfo{author}{\bibfnamefont{A.~M.} \bibnamefont{{Ozorio de Almeida}}},
  \bibinfo{author}{\bibfnamefont{I.}~\bibnamefont{{Dami{\~a}o Soares}}},
  \bibnamefont{and} \bibinfo{author}{\bibfnamefont{E.~V.}
  \bibnamefont{{Tonini}}}, \bibinfo{journal}{Phys. Rev. D}
  \textbf{\bibinfo{volume}{65}}, \bibinfo{pages}{083511}
  (\bibinfo{year}{2002}).

\bibitem[{\citenamefont{{Soares} and {Stuchi}}(2005)}]{TypeIXChaoticScatterer3}
\bibinfo{author}{\bibfnamefont{I.~D.} \bibnamefont{{Soares}}} \bibnamefont{and}
  \bibinfo{author}{\bibfnamefont{T.~J.} \bibnamefont{{Stuchi}}},
  \bibinfo{journal}{Phys. Rev. D} \textbf{\bibinfo{volume}{72}},
  \bibinfo{pages}{083516} (\bibinfo{year}{2005}).

\bibitem[{\citenamefont{{Heinzle} et~al.}(2005)\citenamefont{{Heinzle},
  {R{\"o}hr}, and {Uggla}}}]{TypeIXChaoticScatterer4}
\bibinfo{author}{\bibfnamefont{J.~M.} \bibnamefont{{Heinzle}}},
  \bibinfo{author}{\bibfnamefont{N.}~\bibnamefont{{R{\"o}hr}}},
  \bibnamefont{and} \bibinfo{author}{\bibfnamefont{C.}~\bibnamefont{{Uggla}}},
  \bibinfo{journal}{Phys. Rev. D} \textbf{\bibinfo{volume}{71}},
  \bibinfo{pages}{083506} (\bibinfo{year}{2005}).

\bibitem[{\citenamefont{{Heinzle} et~al.}(2006)\citenamefont{{Heinzle},
  {R{\"o}hr}, and {Uggla}}}]{TypeIXChaoticScatterer5}
\bibinfo{author}{\bibfnamefont{J.~M.} \bibnamefont{{Heinzle}}},
  \bibinfo{author}{\bibfnamefont{N.}~\bibnamefont{{R{\"o}hr}}},
  \bibnamefont{and} \bibinfo{author}{\bibfnamefont{C.}~\bibnamefont{{Uggla}}},
  \bibinfo{journal}{Phys. Rev. D} \textbf{\bibinfo{volume}{74}},
  \bibinfo{pages}{061502} (\bibinfo{year}{2006}).

\end{thebibliography}
\end{document}